\documentclass[12pt,letterpaper]{article}

\usepackage[includeheadfoot,
            marginratio={1:1,2:3}, 
            width=412pt,
            height=688pt,]{geometry}
            \usepackage{tikz}
\usepackage[compat=1.0.0-rc1.1]{tikz-feynman}
\usepackage[utf8]{inputenc}
\usepackage{amsmath}
\usepackage{amsfonts}
\usepackage{amsthm}
\usepackage{amssymb}
\usepackage{graphicx}
\usepackage{booktabs}
\usepackage{empheq}
\usepackage{paralist}
\usepackage{cite}
\usepackage[hidelinks]{hyperref}
\usepackage{xcolor}
\usepackage{tensor}
\usepackage{adjustbox}
\usepackage{subcaption}
\usepackage{braket}
\usepackage{tikz}
\usepackage{tikz-cd}
\usepackage{yfonts}
\usepackage{enumitem}
\usepackage{physics}
\usepackage{mdframed}

\usetikzlibrary{decorations.pathreplacing,calc}

\numberwithin{equation}{section}
\numberwithin{table}{section}

\newcommand{\nc}{\newcommand}
\nc{\lb}{\llbracket}
\nc{\rb}{\rrbracket}
\nc{\gl}{\llbracket}
\nc{\gr}{\rrbracket}
\nc{\bbR}{\mathbb{R}}
\nc{\bbC}{\mathbb{C}}
\nc{\bbZ}{\mathbb{Z}}
\nc{\cO}{\mathcal{O}}
\nc{\cS}{\mathcal{S}}
\nc{\cM}{\mathcal{M}}
\nc{\cT}{\mathcal{T}}
\nc{\cX}{\mathcal{X}}
\nc{\cQ}{\mathcal{Q}}
\nc{\cD}{\mathcal{D}}
\nc{\cC}{\mathcal{C}}
\nc{\cG}{\mathcal{G}}
\nc{\cF}{\mathcal{F}}
\nc{\cI}{\mathcal{I}}
\nc{\pd}{\partial}
\nc{\la}{\lambda}
\newcommand\beq{\begin{equation}}
\newcommand\eeq{\end{equation}}
\nc{\del}{\partial}
\nc{\tri}{\hspace{-31pt}\vartriangle\hspace{-31pt}}
\nc{\blacktri}{\blacktriangle}
\nc{\eq}[1]{\begin{equation}
                     \begin{split} #1 \end{split}
                     \end{equation}}
\nc{\ul}{\underline}
\nc{\ov}{\overline}
\nc{\fa}{\hat}
\nc{\fb}{\MakeUppercase}
\nc{\fc}{\tilde }
\nc{\Lie}{{\cal L}} 
\nc{\lambdabar}{{\mkern0.75mu\mathchar '26\mkern -9.75mu\lambda}}

\renewcommand{\P}{\mathbb{P}}
\newcommand{\R}{\mathbb{R}}

\newcommand{\C}{\mathbb{C}}

\renewcommand{\L}{\mathcal{L}}
\renewcommand{\H}{\mathcal{H}}

\renewcommand{\O}{\mathcal{O}}

\newcommand{\A}{\mathcal{A}}

\newcommand{\N}{\mathbb{N}}
\newcommand{\Z}{\mathbb{Z}}
\newcommand{\M}{\mathcal{M}}
\newcommand{\D}{\mathcal{D}}
\newcommand{\E}{\mathcal{E}}
\newcommand{\I}{\mathcal{I}}

\newcommand{\im}{\mathrm{im}}
\newcommand{\Hom}{\mathrm{Hom}}

\newmuskip\pFqmuskip

\newcommand*\pFq[7][8]{
  \begingroup 
  \pFqmuskip=#1mu\relax
  \mathchardef\normalcomma=\mathcode`,
  \mathcode`\,=\string"8000
  \begingroup\lccode`\~=`\,
  \lowercase{\endgroup\let~}\pFqcomma
  {}_{#2}{#3}_{#4}{\left[\left.\genfrac..{0pt}{}{#5}{#6}\right|#7\right]}
  \endgroup
}
\newcommand{\pFqcomma}{{\normalcomma}\mskip\pFqmuskip}

\allowdisplaybreaks[2]

\tikzfeynmanset{
bulkblob/.style={
/tikz/shape=circle,
/tikz/draw=green,
/tikz/pattern=north west lines,
/tikz/pattern color=green,
/tikz/minimum size = 1.4 cm,
}
}
\tikzfeynmanset{
boundarydot/.style={
/tikzfeynman/dot,
/tikz/color=blue,
}
}
\tikzfeynmanset{
bulkdot/.style={
/tikzfeynman/dot,
/tikz/color=green,
}
}

\begin{document}

\vspace*{0.8cm}

\begin{center}
{\Large
Reductions of GKZ Systems and} \\[.3cm]
{\Large Applications to Cosmological Correlators}

\vspace{.6cm}
\end{center}

\begin{center}
 Thomas W.~Grimm and Arno~Hoefnagels
\end{center}

\vspace{.5cm}
\begin{center} 
\vspace{0.5cm} 
\emph{
Institute for Theoretical Physics, Utrecht University,\\ 
Princetonplein 5, 3584 CC Utrecht, 
The Netherlands } \\
   
\vspace{0.2cm}

\vspace{0.3cm}
\end{center}

\vspace{01cm}


\begin{abstract}
\noindent
A powerful approach to computing Feynman integrals or cosmological correlators is to consider them as solution to systems of differential equations. Often these can be chosen to be Gelfand-Kapranov-Zelevinsky (GKZ) systems. However, their naive construction 
introduces a significant amount of unnecessary complexity. In this paper we present an algorithm which allows for reducing these GKZ systems to smaller subsystems if a parameter associated to the GKZ systems is resonant. These simpler subsystems can then be solved separately resulting in solutions for the full system. The algorithm makes it possible to check when reductions happen and allows for finding the associated simpler solutions. While originating in the mathematical theory of D-modules analyzed via exact sequences of  Euler-Koszul homologies, the algorithm can be used without knowledge of this framework.
We motivate the need for such reduction techniques by considering cosmological correlators on an FRW space-time and solve the tree-level single-exchange correlator in this way. It turns out that this integral exemplifies an interesting relation between locality and the reduction of the differential equations.

\end{abstract}

\clearpage

\tableofcontents


\newpage

\parskip=.2cm
\section{Introduction}

In perturbative quantum field theory, physical amplitudes can be obtained as a finite sum of Feynman integrals. Although these integrals are ordinary integrals, they are often notoriously difficult to evaluate as functions of the kinematic variables and parameters of the theory. However, in many cases, the evaluation can be greatly simplified by interpreting the Feynman integrals as solutions to a particular set of differential equations \cite{anastasiou_higgs_2002,anastasiou_dilepton_2003,anastasiou_automatic_2004,argeri_feynman_2007,cachazo_sharpening_2008,arkani-hamed_allloop_2011,henn_multiloop_2013,gehrmann_qcd_2014,henn_lectures_2015,vanhove_differential_2021,armadillo_evaluation_2023,giroux_looploop_2023,kreimer_bananas_2023,henn_first_2023,gorges_procedure_2023,hidding_feynman_2023,dlapa_algorithmic_2023,britto_generalized_2023,ananthanarayan_method_2023,he_symbology_2023,marzucca_recent_2024,jiang_symbol_2024,he_symbology_2024,fevola_landau_2024,frellesvig_epsilonfactorised_2024,henn_dmodule_2024,calisto_learning_2024}. The choice of differential equations varies significantly, as do the methods used to derive them. For instance, when considering a specific integral, one could use integration by parts identities of the integral \cite{tkachov_theorem_1981,chetyrkin_integration_1981,kotikov_differential_1991,remiddi_differential_1997,gehrmann_differential_2000,artico_integrationbyparts_2024}, or alternatively,  interpret the integral geometrically \cite{bloch_feynman_2015,bloch_local_2017,bourjaily_traintracks_2018,bourjaily_bounded_2019,bourjaily_embedding_2020,klemm_lloop_2020,bonisch_analytic_2021,bonisch_feynman_2022,gasparotto_cohomology_2023,lairez_algorithms_2023,delacruz_algorithm_2024}.

One particular setting in which this approach was used recently is in the study of cosmological correlators \cite{weinberg_quantum_2005,arkani-hamed_cosmological_2015,baumann_bootstrapping_2020,baumann_snowmass_2022,baumann_linking_2022,jazayeri_cosmological_2022,lee_leading_2023,arkani-hamed_differential_2023,arkani-hamed_kinematic_2023,benincasa_geometry_2024,benincasa_asymptotic_2024,fan_cosmological_2024,xianyu_inflation_2024,de_cosmology_2024,aoki_cosmological_2024,alaverdian_difference_2024,chen_systematic_2024,he_differential_2024,baumann_new_2024,melville_sitter_2024,goodhew_cosmological_2024,fevola_algebraic_2024,baumann_kinematic_2024}. These equal time correlators are a natural observable in cosmology where, for example, they can be used to study fluctuations that arise during cosmic inflation \cite{guth_inflationary_1981,albrecht_cosmology_1982,guth_fluctuations_1982,hawking_development_1982,linde_new_1982,starobinsky_dynamics_1982,bardeen_spontaneous_1983,spergel_cmb_1997,hu_distinguishing_1997,dodelson_coherent_2003,mcfadden_holography_2010,assassi_soft_2012,mata_cmb_2013,pajer_boostless_2020,gomez_cosmological_2021,jazayeri_locality_2021,sleight_ds_2021,achucarro_inflation_2022,pimentel_boostless_2022,pinol_cosmological_2023,jazayeri_shapes_2023,chowdhury_subtle_2024,stefanyszyn_there_2024,werth_cosmological_2024}. Since these correlators only depend on the spatial coordinates, they admit the special property that their Fourier transforms satisfy differential equations in the spatial momenta alone. This indicates that the formulation is initially time-independent and that cosmological time could be replaced by alternative measures such as kinematic flow \cite{arkani-hamed_kinematic_2023} or the complexity of the correlators \cite{grimm_structure_2024}. Interestingly, the solutions to these differential equations often turn out to be simpler than the intermediate calculations would suggest. In this paper, we show that this simplification arises from a more general phenomenon -- the reducibility of the differential equation system -- which applies not only to general Feynman integrals and geometric period integrals but also to many other settings.

At first, a notion of simplicity can be somewhat nebulous and open to multiple interpretations. Therefore, it is essential to clarify the types of simplifications we will be considering. Since our focus will be on studying functions through the differential equations they satisfy, a natural measure of simplicity is the number of linearly independent solutions to these differential equations. Generally, a lower-dimensional solution space corresponds to a simpler system to solve and thus should yield simpler functions. Specifically, we will examine cases where some solutions to the complete set of differential equations can also be obtained by solving a different, more restricted set of differential equations. Because this second set is more constrained, it should have fewer linearly independent solutions and be easier to solve.

To illustrate this, let us consider a homogeneous differential equation in one variable. For a general ordinary differential operator $P$ with rational coefficients, reducibility is equivalent to a decomposition $P=Q R$. Here $Q$ and $R$ are also differential operators with rational coefficients with orders strictly lower than the order of $P$. This implies that there is a subset of solutions to
\begin{equation}\label{eq:pf=0}
    Pf(x)=0
\end{equation}
given by the solutions to
\begin{equation}\label{eq:rf=0}
    Rf(x)=0\, .
\end{equation}
Since $R$ is of strictly lower order than $P$, equation~\eqref{eq:rf=0} has fewer solutions than equation~\eqref{eq:pf=0}. Because of this, one might call these solutions simpler or less complex than originally expected. It turns out that this reduction in complexity can be indeed quantified with the measure suggested in \cite{grimm_structure_2024} as we will demonstrate in \cite{futureworkGHV}.

These observations can be extended to systems of differential equations involving multiple variables. In this context, one can again consider the number of linearly independent solutions to the entire system. The concept of reducibility in a differential system can be described as follows:
\begin{center} \textit{A set of differential equations is reducible if there exists a subset of solutions that is annihilated by additional differential operators.} 
\end{center}
These additional operators impose further constraints, resulting in a solution space of lower dimensionality. The usefulness of reducibility lies in the fact that, due to these extra constraints, finding solutions to this second set of differential equations is almost always easier than solving the entire system at once. Moreover, there may be multiple subsystems of a given set of differential equations, each associated with a partial solution basis. In certain cases, such as the example we will explore in this paper, these partial solution bases can be combined to form a complete solution basis for the full set of differential equations.

In this paper we will study the reducibility of a particular system of differential equations, originally due to Gelfand, Kapranov, and Zelevinsky \cite{gelfand_generalized_1990,gelfand_hypergeometric_1991,gelfand_discriminants_1994}. This set of differential equations, which we will refer to as the \textit{GKZ system} \cite{saito_Grobner_2000,nasrollahpoursamami_periods_2016,delacruz_feynman_2019,feng_gkzhypergeometric_2020,klausen_hypergeometric_2020,klausen_kinematic_2022,klausen_hypergeometric_2023a,ananthanarayan_feyngkz_2023,blumlein_hypergeometric_2021,blumlein_hypergeometric_2023,chestnov_restrictions_2023,caloro_ahypergeometric_2023}, can be readily derived for any integral over polynomials raised to complex powers. This means that this framework is quite general and encompasses, for example, all Feynman integrals, cosmological correlators, period integrals and many other classes of integrals. Furthermore, the reducibility of a GKZ system is understood completely and encoded in the complex powers in the integrand \cite{saito_Irreducible_2011,beukers_Irreducibility_2011,schulze_resonance_2012}. Originally these results have been phrased in a mathematical, technically involved language. One aim of this paper is to translate these results into a more accessible form, making them usable without requiring in-depth knowledge of the underlying mathematics.

One complication in the mathematical approach to the reducibility of  GKZ systems, as described in \cite{saito_Irreducible_2011,beukers_Irreducibility_2011,schulze_resonance_2012}, is the reliance on $\D$-modules. This perspective can be quite abstract and may obscure the practical aspects of reducibility, which can be more intuitively understood in terms of partial solution bases. In this work, we demonstrate that the framework developed in \cite{schulze_resonance_2012} can be effectively used to explicitly solve for these partial solution bases. We provide methods to directly obtain the differential operators, which we term \textit{reduction operators}, associated with these partial solution bases thereby offering a concrete algorithm to carry out the reduction of as given GKZ system.

To illustrate the reduction techniques, we will use a specific cosmological correlator as an example. The associated integral, that we will call the single exchange integral, arises from a particular 4-point diagram of a toy model introduced in \cite{arkani-hamed_cosmological_2015}.
This integral has been previously solved using various methods in \cite{arkani-hamed_cosmological_2015,arkani-hamed_differential_2023,arkani-hamed_kinematic_2023}. However, because it clearly demonstrates the simplification process, it will serve as our benchmark example for the techniques presented in this paper.

In the final part of this work, we make the intriguing observation that when applying the reduction operators to the single exchange integral, the resulting differential equations precisely match the conditions imposed by the locality of the underlying theory. As noted in \cite{arkani-hamed_differential_2023}, this locality is connected to having twists in the integral that are defined by non-integer-valued, complex exponents in the integrand. Interestingly, these twists also determine the reducibility of the GKZ system, suggesting a deeper link between locality and reducibility.
This observation can be extended beyond the specific example discussed. While \cite{arkani-hamed_differential_2023} primarily focused on the locality conditions for the single exchange integral, the reducibility of GKZ systems can be examined in a more general context. We believe that further exploration of reducibility in cosmological correlators could lead to a more comprehensive understanding of how these simplifications emerge from locality.

This paper is structured as follows. We begin in section~\ref{sec:ReviewCC} by introducing the model that gives rise to the single exchange integral, the main example we will study in this paper. Furthermore, we briefly discuss how locality is encoded in this integral, setting the stage for the connections with reducability. In section~\ref{sec:gkzsystems} we provide a short introduction to GKZ systems and how these can be obtained for general sets of integrals. As an example, we determine the GKZ system associated to the single exchange integral. The main results of this paper are contained in section~\ref{sec:reductions}. We begin in section~\ref{ssec:reductionoperators} with a description of the two types of subsystems that can exist in a reducible GKZ system. This characterization is followed by an algorithm for obtaining the reduction operators. When these operators are found they can immediately be used to obtain partial solution bases. However, it can be useful to classify exactly when these operators exist and lead to non-trivial subsystems as we explain in section~\ref{ssec:reducible}.

Finally, we apply this machinery to the single exchange integral in section~\ref{sec:singexreducs}. We follow the structure of section~\ref{sec:reductions} and begin by obtaining the different reduction operators in section~\ref{ssec:reduction_operators_ex}. Afterwards, we use the reduction operators to obtain the solutions to the single exchange system in sections~\ref{ssec:obtaining_solutions}, \ref{ssec:explicitsols} and~\ref{ssec:qimoving}, showcasing two complementary ways the reduction operators can be used. In section~\ref{ssec:physics} we show how locality is encoded in the reduction operators, and speculate how these reduction operators can be used to study locality in more general terms. 

We also include two appendices in this paper. Appendix~\ref{ap:GKZ} introduces some useful properties of GKZ systems, while appendix~\ref{ap:derivation} contains the arguments for the general statements made in the main text. The latter appendix also explains how our results follow from the statements made in the mathematical literature.

\section{Motivation -- cosmological correlators}\label{sec:ReviewCC}

In this section, we introduce a primary use for the general reduction algorithm for GKZ systems described in this work: the computation of correlators in a cosomolgical spacetime. 
This physical setting will also provide us with a concrete example to which we will apply the reduction algorithm in the course of the paper. 
The following discussion will largely follow the recent exposition \cite{arkani-hamed_differential_2023}. The toy example that we will introduce was first described in \cite{arkani-hamed_cosmological_2015}.

\paragraph{Correlators and the wave function of the universe.} 
In cosmology one is naturally led to study  correlation functions that are only determined by an initial state. 
In fact, instead of the usual $\bra{in}\ket{out}$-correlations that one uses to calculate a flat space S-matrix, one is investigating the in-in correlations that take the form \cite{baumann_snowmass_2022,weinberg_quantum_2005}
\begin{equation}
    \langle\; \prod_i \O_i(\Vec{x}_i,\eta_i) \;\rangle = \bra{in}\prod_i \O_i(\Vec{x}_i,\eta_i)\ket{in}\ ,
\end{equation}
where $\Vec{x}$ denotes the spatial coordinates and $\eta$ is the time coordinate in the usual FRW metric $ds^2=a(\eta)^2(d\eta^2 + d\Vec{x}^2)$. When studying inflation  one fixes the correlators to the time-slice $\eta_i=0$. Furthermore, one considers the in-state to be the Bunch-Davies vacuum at $\eta_i \rightarrow -\infty$. One might then try to calculate these correlators directly in perturbation theory. However, this will involve implementing a mixture of time-ordering and reverse-time-ordering which can make these calculations quite subtle \cite{baumann_snowmass_2022}. 

Another useful approach is to introduce the wave-function of the universe \cite{anninos_Latetime_2015,arkani-hamed_cosmological_2017}. This is defined as the wave-functional $\Psi[\varphi]=\braket{\varphi}{0}$, where $\ket{\varphi}$ is an eigenstate of the field operator $\hat{\phi}(\Vec{x},\eta)$ with $\hat{\phi}(\Vec{x},0)\ket{\varphi}=\varphi(\Vec{x})\ket{\varphi}$. The use of this wave-form becomes clear when one tries to calculate equal time correlation functions of the fields $\varphi$, since these can be expressed as
\begin{equation}
    \langle \;\prod_i \varphi(\Vec{x}_i)\;\rangle = \frac{\int \D \varphi\;  \vert \Psi[\varphi]\vert^2  \prod_i \varphi(\Vec{x}_i)}{\int \D \varphi\;  \vert \Psi[\varphi]\vert^2} \, .
\end{equation}
Therefore, finding the wave-function of the universe leads to an alternative way of calculating correlators.

As is often the case, it is useful to perform an expansion of this wave-function. To be precise, we will expand its logarithm as
\begin{equation}
    \log(\Psi[\varphi])=\sum_{n=0}^\infty \frac{1}{n!}\int \prod_{i=1}^n \left( d^3\Vec{x}_i \;\varphi(\Vec{x}_i)\right)\; \Psi_n(\Vec{x}_1,\cdots,\Vec{x}_n)
\end{equation}
and we are interested in finding the functions $\Psi_n(\Vec{x}_1,\cdots,\Vec{x}_n)$. It turns out to be useful to consider their Fourier transforms $\Psi_n(\Vec{k})$ and factor out the delta function to obtain
\begin{equation}
    \Psi_n(\Vec{k}_1,\cdots,\Vec{k}_n)=\delta^3\left( \sum_{i=1}^n \Vec{k}_i \right) \psi_n(\Vec{k}_1,\cdots,\Vec{k}_n) \, .
\end{equation}
The functions $\psi_n(\Vec{k}_1,\cdots,\Vec{k}_n)$ are called the \textit{wave-function coefficients}.

\paragraph{The toy model.} To do this, we will introduce a toy model which has the particular property that it is possible to find correlators for a variety of space-times simultaneously. We will consider a conformally coupled scalar in an FRW space-time with generic polynomial interactions. This model was introduced in \cite{arkani-hamed_cosmological_2017} and we will mostly follow the exposition  of \cite{arkani-hamed_cosmological_2017,arkani-hamed_differential_2023}. 
The action for this model can be written as 
\begin{equation}
    S=\int d^3\vec{x} d\eta \; \left( \frac{1}{2}(\partial \phi)^2 -\sum_{n\geq 3} \frac{\lambda_n(\eta)}{n!} \phi^n \right)\, ,
\end{equation}
where the $\lambda_n$ are time-dependent coupling constants $
    \lambda_n(\eta)=\lambda_{n,0} \left(a(\eta) \right)^{4-n}$
and $a(\eta)$ is the pre-factor of the FRW metric. The case we will consider consist of space-times which have 
\begin{equation}
    a(\eta)= \left(\frac{\eta}{\eta_0}\right)^{-(1+\epsilon)}\, ,
\end{equation}
where $\eta_0$ and $\epsilon$ will be kept arbitrary for now. 

\begin{figure} 
\centering

\begin{tikzpicture}
    \begin{feynman}[large]
        \vertex (topleft);
        \vertex [right=2 cm of topleft,boundarydot] (topcenter1) {};
        \vertex [right=2 cm of topcenter1,boundarydot] (topcenter2) {};
        \vertex [right=1 cm of topcenter2,boundarydot] (topcenter);
        \vertex [below=2 cm of topcenter,bulkblob] (diagram) {};
        \vertex [right=.4 cm of topcenter] (topcenter3) {\quad \quad\quad};
        \vertex [right=2.6 cm of topcenter3,boundarydot] (topcenter4) {};
        \vertex [right =2 cm of topcenter4] (topright) {\color{blue} $\eta=0$};

        \diagram*  {
          (topleft) --[very thick,blue] (topcenter1) --[very thick,blue] (topcenter2) --[very thick,blue] (topcenter3) --[very thick,blue] (topcenter4) --[very thick,blue] (topright);
          (diagram) --[thick,opacity=1] (topcenter1);
          (diagram) --[thick,opacity=1]  (topcenter2);
          (diagram) --[thick,opacity=1] (topcenter4);
        };
        \vertex [below=1 cm of topcenter2] (k2) {$\Vec{k}_2$};
        \vertex[left=1.5 cm of k2] (k1) {$\Vec{k}_1$};
        \vertex[right=1.8 cm of k2] (dots) {$\cdots$};
        \vertex[right=2.2 cm of dots] (kn) {$\Vec{k}_n$};
    \end{feynman}
\end{tikzpicture}
\caption{The general in-in diagram corresponding to $n$ external 
particles.}\label{fig:generalinin}
\end{figure}
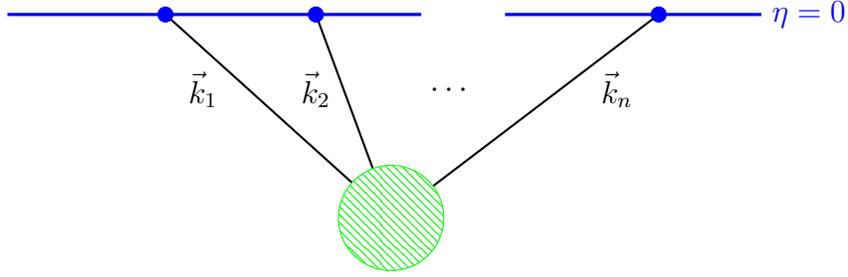

One can calculate the $n$-th wave-function coefficient $\psi_n$ using a diagrammatic approach. One draws all the possible diagrams with $n$ external particles ending on a single time-slice $\eta=0$, as is drawn in figure~\ref{fig:generalinin}. 
The crucial point in this evaluation is that the time of insertion is arbitrary and therefore needs to be integrated over. This means that every vertex in the Feynman diagram introduces an integral of the form
\begin{equation}
    \int_{-\infty}^0 d\eta_v\; \lambda_{n,0} \left( \frac{\eta}{\eta_0} \right)^{(n-4)(1+\epsilon)}\, .
\end{equation}
Therefore, even tree diagrams will contain integrals and are non-trivial. With this in mind one obtains the Feynman rules summarized in \cite{arkani-hamed_cosmological_2017} and can derive an integral expression for any in-in diagram of interest. 

A useful trick is to introduce the frequency representations of the $\lambda_n$, defined by
\begin{equation}
     \lambda_n(\eta)=\int_0^\infty d\omega \;\tilde{\lambda}_n(\omega) e^{i\omega \eta}
\end{equation}
and insert this for every vertex factor. After this insertion the Feynman diagram for a graph $G$ can be rewritten as an integral over
the flat space Feynman wavefunction coefficient for the graph $G$.
The advantage of considering this particular representation of the Feynman integral is that, for tree-level diagrams, the flat space wave-function coefficients will be rational functions and can be obtained combinatorially \cite{arkani-hamed_cosmological_2015,arkani-hamed_differential_2023}. With this representation, we will now introduce the four particle single-exchange diagram.

\begin{figure}[ht]
\centering
\begin{tikzpicture}
    \begin{feynman}[large]
        \vertex (topleft);
        \vertex [right=2 cm of topleft,boundarydot] (topcenter1) {};
        \vertex [right=2 cm of topcenter1,boundarydot] (topcenter2) {};
        \vertex [right=1 cm of topcenter2,boundarydot] (topcenter);
        \vertex [below=2 cm of topcenter] (midcenter) ;
        \vertex[left=2cm of midcenter,bulkdot] (leftdiagram) {};
        \vertex[right=2cm of midcenter,bulkdot] (rightdiagram){};
        \vertex [right=1 cm of topcenter,boundarydot] (topcenter3) {};
        \vertex [right=2 cm of topcenter3,boundarydot] (topcenter4) {};
        \vertex [right =2 cm of topcenter4] (topright) {\color{blue} $\eta=0$};

        \diagram*  {
          (topleft) --[very thick,blue] (topcenter1) --[very thick,blue] (topcenter2) --[very thick,blue] (topcenter3) --[very thick,blue] (topcenter4) --[very thick,blue] (topright);
          (leftdiagram) --[thick,opacity=1] (topcenter1);
          (leftdiagram) --[thick,opacity=1]  (topcenter2);
          (rightdiagram) --[thick,opacity=1] (topcenter3);
          (rightdiagram) --[thick,opacity=1] (topcenter4);
          (rightdiagram) --[very thick,green] (leftdiagram);
        };
        \vertex [below=1 cm of topcenter2] (k2) {$\Vec{k}_2$};
        \vertex[below=1cm of topcenter1] (k1) {$\Vec{k}_1$};
        \vertex[below=1cm of topcenter3] (k3) {$\Vec{k}_3$};
        \vertex[below=1cm of topcenter4] (k4) {$\Vec{k}_4$};
        \vertex[below=1cm of leftdiagram] (X1) {\textbf{$X_1$}};
        \vertex[below=1cm of rightdiagram] (X2) {\textbf{$X_2$}};
        \vertex[above=0cm of midcenter] (Y) {\textbf{\color{green} $Y$}};
    \end{feynman}
\end{tikzpicture}
\caption{The 4-point single exchange diagram. Here $X_1=\vert \vec{k}_1\vert+\vert\vec{k}_2\vert$, $X_2=\vert \vec{k}_3\vert+\vert\vec{k}_4\vert$ and $Y=\vert \vec{k}_1+\vec{k}_2\vert=\vert \vec{k}_3+\vec{k}_4\vert$.  } \label{fig:singexchange}
\end{figure}
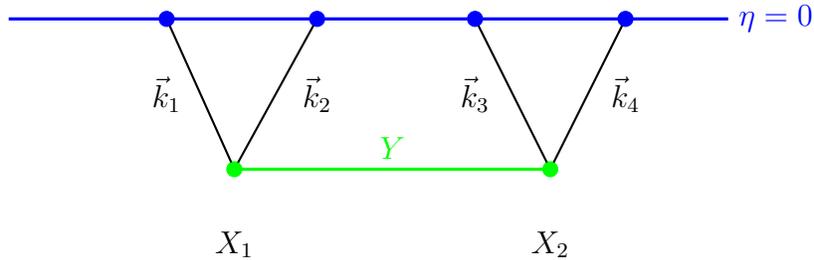

\paragraph{The four particle single exchange integral.} The diagram we will consider is shown in figure~\ref{fig:singexchange} and involves four external lines, as well as a single propagator. Applying the Feynman rules discussed in \cite{arkani-hamed_cosmological_2017} results in the integral
\begin{equation}\label{eq:propaexchange}
\begin{split}
  -\lambda_{3,0}^2  \int_0^{\infty}\omega_1^\epsilon d\omega_1 \int_0^{\infty} \omega_2^\epsilon d\omega_2 \int_{-\infty}^0 d\eta_1 \int_{-\infty}^0 d\eta_2\; e^{i \eta_1(X_1+\omega_1)+i\eta_2(X_2+\omega_2)}G(Y,\eta_1,\eta_2)\, ,
\end{split}
\end{equation}
where $\lambda_{3,0}$ is the constant part of the three-point vertex and $G(Y,\eta_1,\eta_2)$ is the bulk-to-bulk propagator given 
by 
\beq \label{eq:prop-explicit}
  G(Y,\eta_1,\eta_2)= \frac{1}{2Y}\Big(e^{-iY(\eta_1-\eta_2)}\theta(\eta_1-\eta_2)+e^{iY(\eta_2-\eta_1)}\theta(\eta_2-\eta_1) -e^{iY(\eta_1+\eta_2)} \Big)\,. 
\eeq
Inserting this propagator and performing the $\eta_i$ integrals, one obtains
\begin{equation}\label{eq:singleexchange}
  \boxed{\rule[-.5cm]{0cm}{1.3cm} \quad I(X_1,X_2,Y,\epsilon)\coloneqq \int_0^\infty \frac{ \lambda_{3,0}^2(\omega_1 \omega_2)^\epsilon d\omega_1 d\omega_2}{(\omega_1+X_1+Y)(\omega_2+X_2+Y)(\omega_1+\omega_2+X_1+X_2)}\, .\quad }
\end{equation}
This is the integral we will consider throughout the text and use to illustrate the reduction algorithm.

\paragraph{Comments on locality.} The single-exchange integral \eqref{eq:singleexchange} has special properties that are linked to the locality of the underlying theory \cite{arkani-hamed_differential_2023}. To make this more precise we replace the propagator $G(Y,\eta_1,\eta_2)$ by $-i \delta(\eta_1-\eta_2)$, which 
is equivalent to collapsing the propagator to a point.
The integrations over the $\omega_i$ can then be performed explicitly and result in 
\begin{equation}\label{eq:icont1}
   I_\mathrm{contr}= -2^{2(\epsilon +1)}\sqrt{\pi} \text{csc}(\pi \epsilon) \Gamma\big(-\epsilon -\tfrac{1}{2} \big) \Gamma(\epsilon+1) (X_2+X_1)^{2 \epsilon +1}\, ,
\end{equation}
where we have defined $I_\mathrm{contr}$ to be the contracted integral. 
Alternatively, we can use the identity
\begin{equation}
    (\partial_{\eta_1}^2+Y^2)G(Y,\eta_1,\eta_2)=-i \delta(\eta_1-\eta_2)
\end{equation}
in the integrand and integrate by parts twice.
After integration over $\eta_1$ and $\eta_2$ this leads to the alternative expression
\begin{equation}\label{eq:icont2}
   I_\mathrm{contr}= \int_0^\infty \frac{ \lambda_{3,0}^2(\omega_1^2+2\omega_1 X_1+X_1^2-Y^2) (\omega_1 \omega_2)^\epsilon d\omega_1 d\omega_2}{(\omega_1+X_1+Y)(\omega_2+X_2+Y)(\omega_1+\omega_2+X_1+X_2)}\ .
\end{equation}
The crucial observation is that it is now possible to translate the equality of equations~\eqref{eq:icont1} and~\eqref{eq:icont2} to a differential equation on $I$ by repeatedly applying the integration by parts identity
\begin{equation}\begin{split}
    &\frac{\partial}{\partial X_1} \int_0^\infty \frac{ \omega_1^\alpha \omega_2^\epsilon d\omega_1 d\omega_2}{(\omega_1+X_1+Y)(\omega_2+X_2+Y)(\omega_1+\omega_2+X_1+X_2)}\\
    =& -\alpha \int_0^\infty \frac{ \omega_1^{\alpha-1}\omega_2^\epsilon d\omega_1 d\omega_2}{(\omega_1+X_1+Y)(\omega_2+X_2+Y)(\omega_1+\omega_2+X_1+X_2)}
    \end{split}
\end{equation}
to equation~\eqref{eq:icont2}. This leads to the differential equation
\begin{equation}\label{eq:locality}
    \left(X_1^2-Y^2\right) \frac{\partial^2 I}{\partial X_1^2}+2 X_1(1-\epsilon) \frac{\partial I}{\partial X_1}- \epsilon (1-\epsilon)I=\frac{\partial^2 I_{\mathrm{contr}}}{\partial X_1^2}\, .
\end{equation}
This differential equation relates the single exchange integral with the contracted integral $I_\mathrm{contr}$ that we evaluated in equation~\eqref{eq:icont1}. Therefore, one can loosely think of the differential operator on the left hand-side as contracting the propagator when acting on $I$. Note that a similar derivation can be done for $X_2$, also leading to a second order differential equation for $I$. 

Interestingly, the differential operator on the left hand side only involves $X_1$ and derivatives with respect to $X_1$. As we will see below, $I$ solves a pair of second order differential equations which mixes $X_1$ and $X_2$, as is generally the case. However, due to locality one finds that these differential equation must split, a highly non-trivial property. It turns out that this splitting is deeply connected to the reducibility of the associated GKZ system, a fact which we will explore further in section~\ref{ssec:physics}. Before this, we will need to discuss the GKZ system associated to this integral.

\section{GKZ systems and the single exchange integral}\label{sec:gkzsystems}

As stressed in the introduction, our strategy to solving integrals such as~\eqref{eq:singleexchange} is to determine first a set of differential equations that admit~\eqref{eq:singleexchange} as a solution, and then obtain an explicit expression for~\eqref{eq:singleexchange} by appropriately combining the basis elements of the solution space. The specific set of differential equations that we will consider is the associated GKZ system \cite{gelfand_generalized_1990,gelfand_hypergeometric_1991}. The goal of this section is to describe the structure of such systems. 
We will begin with the general theory of GKZ systems in section~\ref{ssec:gengkz}. These systems of differential equations are obtained by defining an integer matrix $\mathcal{A}$ and a vector $\nu$ associated to a given integral. 
The GKZ system and the data $\mathcal{A}$, $\nu$ associated to the integral~\eqref{eq:singleexchange} will be determined in 
section~\ref{ssec:singexch}.  

\subsection{A brief introduction to GKZ systems}\label{ssec:gengkz}

A GKZ system can be obtained for any integral of the form
\begin{equation}\label{eq:gengkzint}
   I(z;\alpha,\beta)= \int_\Gamma d^n \omega\; \frac{\prod_{i=1}^n \omega_i^{\alpha_i-1}}{\prod_{j=1}^k p_j(z,\omega)^{\beta_j}}\, ,
\end{equation}
where $\Gamma$ is an arbitrary integration cycle, $\alpha_i$ and $\beta_j$ are complex numbers and the $p_j$ are polynomials in the $\omega_i$ with coefficients $z_{j,n}$. To be specific we will take $p_j$ to be written as
\begin{equation}
    p_j(z,\omega)= \sum_m z_{j,m} \prod_{i=1}^n \omega_i^{(a_{j,m})_i}\, ,
\end{equation}
where the $a_{j,m}$ are vectors describing the powers of $x$ in each term of $p_j$. For example, the polynomial
\begin{equation}\label{eq:gkzexamp}
    p_j(z,\omega)=z_{j,1} \;\omega_2+z_{j,2}\;\omega_1^3 \omega_2^2
\end{equation}
results in the vectors $a_{j,1}=(0,1)^T$ and $a_{j,2}=(3,2)^T$, with $T$ being the transpose. The coefficients $z_{j,m}$ will function as the variables that the integral depends on and the differential equations solved by the integral will be differential equations in these variables. 

\paragraph{The matrix defining a GKZ system.}

The differential equations defining a GKZ system are defined from a certain matrix $\A$. These matrices follow from the polynomials $p_j$ as follows. One begins by defining a set of matrices $\A_j$ for each polynomial $p_j$, simply by interpreting the vectors $a_{j,m}$ as column vectors of these matrix. Continuing with the example of equation~\eqref{eq:gkzexamp}, this matrix takes the form
\begin{equation}
    \A_j=\begin{pmatrix}
    0 & 3\\
    1 & 2
    \end{pmatrix}\, .
\end{equation}
One combines the matrices $\A_j$ in a new matrix $\A$ as 
\begin{equation}
    \A\coloneqq \begin{pmatrix} \mathbf{1} & \mathbf{0} &\cdots\\
    \mathbf{0} & \mathbf{1} & \cdots \\
    \vdots&\ddots&\ddots\\
    \A_1 & \A_2 &\cdots 
    \end{pmatrix}\ ,
\end{equation}
where the $\mathbf{1}$ are row vectors with $1$ at every entry and the $\mathbf{0}$ are row vectors of zeroes to fill in the gaps. Adding these zeroes and ones is called homogenization and the full matrix $\A$ is called the homogenized matrix. Combined with the paramaters $\alpha_i$ and $\beta_j$, this determines the GKZ system completely. Again, it is much easier to see this by example where combining the two matrices
\begin{equation}
    \begin{array}{lr}
        \A_1=\begin{pmatrix}
    0 & 3\\
    1 & 2
    \end{pmatrix}\, ,    &     \A_2=\begin{pmatrix}
    2 & 1\\
    0 & 0
    \end{pmatrix}\, ,
    \end{array}
\end{equation}
results in the homogenized matrix
\begin{equation}\label{eq:Aexample}
    \A= \begin{pmatrix}
    1 & 1 & 0 & 0\\
    0 & 0 & 1 & 1\\
    0 & 3 & 2 & 1\\
    1 & 2 & 0 & 0
    \end{pmatrix}\, ,
\end{equation}
and we can identify $\A_1$ and $\A_2$ as being contained in the bottom two rows of $\A$.

\paragraph{Differential equations from the matrix $\A$.}

The differential equations which the integral~\eqref{eq:gengkzint} is a solution to can be obtained from the matrix $\A$. But, before we explain this, it is useful to introduce some simplifying notation. First of all, we will write $N$ for the number of columns of $\A$, and $M$ for the number of rows. Secondly, we will write $a_I$ for the $I$-th column vector of $\A$. Since we can associate a variable $z_{j,m}$ with each column of $\A$, we will rename these to $z_I$ with $I$ being the corresponding column. Furthermore, we will combine the exponent vectors $\alpha$ and $\beta$ from the integral~\eqref{eq:gengkzint} in a single vector $\nu\coloneqq (\beta,\alpha)^T$. We will also use the shorthand
\begin{equation}
    \partial_I\coloneqq \frac{\partial}{\partial z_I}
\end{equation}
for derivatives with respect to $z_I$ and
\begin{equation}\label{eq:thetadef}
    \theta_I\coloneqq z_I \partial_I
\end{equation}
for the associated homogeneous derivative. Finally we will write solutions to the differential equations at a parameter vector $\nu$ as $f(z;\nu)$. With this notation in place we will proceed with obtaining the differential equations.

These differential equations can be separated into two subsets, the toric equations and the Euler equations. Why the integral~\eqref{eq:gengkzint} is a solution of these differential equations is explained in appendix~\ref{ap:difeqs}, here we will just state the differential equations themselves. 

To obtain the toric equations we have to find vectors $u$ and $v$ in $\N^N$ satisfying  
\begin{equation}\label{eq:toricdef}
    \A u=\A v \, .
\end{equation}
Any such $u$ and $v$ give rise to a toric operator $\L_{u,v}$, defined as
\begin{equation}\label{eq:toricopdef}
   \L_{u,v}\coloneqq \prod_{I=1}^N \partial_I^{u_I}-\prod_{I=1}^N\partial_I^{v_I}
\end{equation}
and a toric equation
\begin{equation}\label{eq:gentoriceq}
   \boxed{\rule[-.1cm]{0cm}{.55cm}\quad \L_{u,v}f(z;\nu)=0\ .\quad }
\end{equation}
To find all such $u$ and $v$ one typically uses that, for every vectors $u_1$, $u_2$, $v_1$, and $v_2$ satisfying
\begin{equation}
    \L_{u_1,v_1}f(z;\nu)=\L_{u_2,v_2}f(z;\nu)=0 \, ,
\end{equation}
then their sum must also have $\L_{u_1+u_2,v_1+v_2}f(z;\nu)=0$ since the above equation implies
\begin{equation}
    \prod_{I=1}^N\partial_I^{(v_1)_I+(v_2)_I}f(z;\nu)=\prod_{I=1}^N \partial_I^{(u_1)_I+(u_2)_I}f(z;\nu)\, .
\end{equation}
Therefore, one can find a basis for the integer kernel $\ker_\Z(\A)$ use this to obtain a generating set for the possible combinations of $u$ and $v$.

The second set of differential equations satisfied by our integral are called the Euler equations and are defined as follows. First one takes the differentials $\theta_I$ from equation~\eqref{eq:thetadef} and combines them in a vector
\begin{equation}\label{eq:bigThetadef}
    \Theta\coloneqq (\theta_I)_{1\leq I \leq N}\, .
\end{equation}
One can apply the matrix $\A$ to this vector to obtain the vector
\begin{equation}\label{eq:Edef}
    \E \coloneqq \A \Theta\, .
\end{equation}
The components $\E_J$ of this vector are called Euler operators, where $1\leq J \leq M$. Combined with the parameter vector $\nu$ these give the Euler equations, written as
\begin{equation}\label{eq:eulergen}
    \boxed{\rule[-.1cm]{0cm}{.55cm} \quad(\E_J+\nu_J)f(z;\nu)=0\ , \quad}
\end{equation}
where $1\leq J\leq M$. Note that these are the only equations where $\nu$ plays a role and when considering a single GKZ system one usually considers $\nu$ fixed. Although there are ways of relating solutions at different $\nu$, as we will discuss in section~\ref{sec:reductions}.

\paragraph{Structure of the solutions.}

The original integral~\eqref{eq:gengkzint} will be a particular solution to the above differential equations but usually not the only one. To obtain the integral of interest one first determines a complete basis of solutions $f_{d}(z;\nu)$ to the  differential equations  and afterwards fix the particular coefficients either numerically or by evaluating the integral in specific limits for the $z_I$. Afterwards the integral can then be written as
\begin{equation}
    I(z;\alpha,\beta)=\sum_{d=1}^D c_d(\Gamma;\nu)f_{d}(z;\nu)\, ,
\end{equation}
where $D$ is the dimension of the solution space associated to the GKZ system. Note that $D$ is often also called the rank of the GKZ system.

To find the solutions $f_d$ one can take a few different approaches, of which we want to discuss two in more detail.
One approach is more common in the mathematics literature and results in a complete basis of convergent series expansions \cite{saito_Grobner_2000,cattani_three_2006}. The other approach is more commonly used in physics literature and consists of making a convenient Ansatz that automatically solves the Euler equations \cite{hosono_mirror_1995,hosono_gkzCY_1996,hosono_gkzapp_1996}. This then results in a smaller set if differential equations that can be solved using a variety of methods. In this paper we will mostly use the second method since this requires the least amount of additional theory. The more mathematical approach has some advantages, however, such as giving an easy method of finding the number of solutions as well as always providing a convergent series expansion for each solution. The problem is that these summations can be difficult to evaluate to a more workable form. Since we will mostly use the physics-preferred approach, we leave an explanation of the mathematical approach to appendix~\ref{ap:series} and will now discuss the particular type of Ansatz used. 

The idea of \cite{hosono_mirror_1995,hosono_gkzCY_1996,hosono_gkzapp_1996} is to use that the Euler equations mostly determine the scaling of the solutions, while the toric equations determine their actual form. To see this, we let $u_i$, $v_i$ be $\mathrm{dim}(\ker(\A))$ independent vectors satisfying
\begin{equation}
    \A u_i=\A v_i
\end{equation}
and define associated variables
\begin{equation}\label{eq:sidef}
    s_i\coloneqq \frac{\prod_{I=1}^N z_I^{(u_i)_I}}{\prod_{I=1}^N z_I^{(v_i)_I}}\, ,
\end{equation}
which are known as the \textit{homogeneous variables}. Then, any function of the form
\begin{equation}\label{eq:fpgansatz}
    f(z;\nu)=P(z;\nu)g(s;\nu)
\end{equation}
will have
\begin{equation}
    (\E_J+\nu_J)f(z;\nu)=g(s;\nu)(\E_J+\nu_J)P(z;\nu)\, ,
\end{equation}
where $g(s;\nu)$ denotes that $g(s;\nu)$ can only depend on the variables $s_i$, and not other combinations of the $z_I$.\footnote{Note that $g(s;\nu)$ is also dependent on $\nu$, since the toric equations will mix $P$ and $g$.} Therefore, given any particular solution $P$ to 
\begin{equation}
    (\E_J+\nu_J)P(z;\nu)=0\, ,
\end{equation}
it is possible to make an Ansatz of the form~\eqref{eq:fpgansatz} which is guaranteed to  satisfy the Euler equations. This is especially useful since the Euler equations can be easily solved by considering their scaling properties. Inserting the Ansatz into the toric equations results in a system of partial differential equations for $g(s;\nu)$. Often this system is too difficult to solve directly. However, in this paper we show that if the underlying GKZ is reducible we can find other differential operators that annihilate particular solutions. In these cases, if a convenient Ansatz for $P(z;\nu)$ is chosen it is actually possible to solve the resulting differential equations for $g(s;\nu)$ as we will discuss in~\ref{sec:reductions}.

Now that we have discussed some general properties of GKZ systems and their solutions, we will turn our attention to the system of interest for this paper: the single exchange integral from equation~\eqref{eq:singleexchange}.

\subsection{The single exchange integral}
\label{ssec:singexch}
Having discussed the general theory of GKZ systems, we are now ready to consider the actual integral from equation~\eqref{eq:singleexchange} and its associated GKZ system. To do this, it is first necessary to cast it into the form of~\eqref{eq:gengkzint} by promoting the coefficients in the polynomials to parameters $z_I$ and the exponents to complex numbers $\alpha_i$ and $\beta_i$. For the single exchange integral this results in the Euler integral
\begin{equation}\label{eq:exchangegkz}
    I_\Gamma(z;\nu)=\int_\Gamma d^2\omega\frac{\omega_1^{\alpha_1-1} \omega_2^{\alpha_2-1}}{(z_1+z_2 \omega_1)^{\beta_1}(z_3 +z_4 \omega_2)^{\beta_2}(z_5 +z_6 \omega_1 +z_7 \omega_2)^{\beta_3}}\ ,
\end{equation}
and the original integral can be recovered by inserting
\begin{equation}\label{eq:physvars}
    \begin{array}{rl}
z&=(X_1+Y,1,X_2+Y,1,X_1+X_2,1,1)\ ,\\
\nu&=(\beta,\alpha)=(1,1,1,1+\epsilon,1+\epsilon)^T\ ,\\
\Gamma&=\R_+^2\, .
    \end{array}
\end{equation} 
Comparing the form of the integral above with the general form of equation~\eqref{eq:gengkzint} one sees that it consists of the three polynomials
\begin{equation}
    \begin{array}{rl}
        p_1(\omega,z) &= z_1+z_2 \omega_1\ , \\
        p_2(\omega,z) &= z_3+z_4 \omega_2 \ ,\\
        p_3(\omega,z) &= z_5+z_6 \omega_1 + z_7 \omega_2 \, .
    \end{array}
\end{equation}
Note that the relabeling of $z_{j,m}\rightarrow z_I$ described in the previous section has already taken place. Continuing with the process described there, we define the matrices
\begin{equation}
    \begin{array}{rlrlrl}
       \A_1  &\coloneqq  \begin{pmatrix}
           0 & 1 \\
           0 & 0
       \end{pmatrix},&
       \A_2  &\coloneqq \begin{pmatrix}
           0 & 0 \\
           0 & 1
       \end{pmatrix},&
        \A_3  &\coloneqq  \begin{pmatrix}
           0 & 1 & 0\\
           0 & 0 & 1
       \end{pmatrix}
    \end{array}
\end{equation}
from the polynomials $p_j$. These matrices can be combined into the matrix
\begin{equation}\label{eq:exchangeA}
 \A =\left(
\begin{array}{ccccccc}
 1 & 1 & 0 & 0 & 0 & 0 & 0 \\
 0 & 0 & 1 & 1 & 0 & 0 & 0 \\
 0 & 0 & 0 & 0 & 1 & 1 & 1 \\
 0 & 1 & 0 & 0 & 0 & 1 & 0 \\
 0 & 0 & 0 & 1 & 0 & 0 & 1 \\
\end{array}
\right)\, ,
\end{equation}
where the bottom two rows correspond to the matrices $\A_1$, $\A_2$ and $\A_3$. From this matrix we will now obtain the differential operators of the GKZ system.

Next we determine the toric operators. Since $\mathrm{dim}(\ker(\A))=2$, there are two linearly independent solutions to
\begin{equation}
    \A u= \A v\, ,
\end{equation}
with $u$ and $v$ in $\N^7$. Here we will consider
\begin{equation}
\begin{array}{cccc}
    u_1= \begin{pmatrix}1\\0\\0\\0\\0\\1\\0\end{pmatrix} ,& v_1= \begin{pmatrix}0\\1\\0\\0\\1\\0\\0\end{pmatrix} ,&u_2= \begin{pmatrix}0\\0\\1\\0\\0\\0\\1\end{pmatrix} ,&v_2= \begin{pmatrix}0\\0\\0\\1\\1\\0\\0\end{pmatrix} ,
\end{array}
\end{equation}
satisfying $\A u_1=\A v_1$ and $\A u_2 =\A v_2$. Using the definition of the toric operators from equation~\eqref{eq:toricopdef}, these result in the toric operators
\begin{equation}\label{eq:exchangetoricop}
    \begin{array}{rl}
         \L_{u_1,v_1}&=\partial_1 \partial_6-\partial_2\partial_5\, ,\\
         \L_{u_2,v_2}&=\partial_3 \partial_7 -\partial_4 \partial_5\, ,
    \end{array}
\end{equation}
and toric equations
\begin{equation}\label{eq:exchangetoric}
    \L_{u_1,v_1}f(z;\nu)=\L_{u_2,v_2}f(z;\nu)=0 \, .
\end{equation}
From these and equation~\eqref{eq:sidef}, the parameters $s_i$ can be obtained, resulting in
\begin{equation}\label{eq:sumvars}
    \begin{split}
        s=\frac{z_1 z_6}{z_2 z_5}\, ,& \quad  t=\frac{z_3 z_7}{z_4z_5}\ ,
    \end{split}
\end{equation}
where we have relabeled $s=s_1$ and $t=s_2$.

The Euler operators are obtained from the matrix $\A$ by calculating $\E=\A\Theta$, and take the form
\begin{equation}\label{eq:exchangeeuler}
    \begin{array}{lll}
       \E_1= \theta_1+\theta_2\, , & \E_2=\theta_3 +\theta_4\, , & \E_3=\theta_5 +\theta_6 +\theta_7\ ,\\
        \E_4=\theta_2+\theta_6 \, , & \E_5=\theta_4+\theta_7 \, . &
    \end{array}
\end{equation}
The Euler equations following from these operators are
\begin{equation}
    (\E_J+\nu_J)f(z;\nu)=0\ ,
\end{equation}
for $1\leq J \leq 5$. These two sets of differential equations describe the complete GKZ system associated to the integral~\eqref{eq:exchangegkz}. 

To obtain the solution basis for this system is somewhat non-trivial. While it is easy to obtain a basis of converging series expansions as described in appendix~\ref{ap:series}, these infinite sums are not easily evaluated, even when specializing the parameters $z_I$ and $\nu$ to the physically interesting values from equation~\eqref{eq:physvars}.\footnote{Using some tricks it is actually possible to perform the summations for this particular system, however we consider this system in this paper because it very nicely illustrates the reduction technique described below.} Alternatively, one can try to make an Ansatz like
\begin{equation}\label{eq:generalansatz}
    f(z;\nu)=\frac{z_3^{\nu_5}z_5^{\nu_4}}{z_1^{\nu_1} z_3^{\nu_2} z_4^{\nu_5}z_5^{\nu_3}z_6^{\nu_4}}g(s,t;\nu)
\end{equation}
such as described in equation~\eqref{eq:fpgansatz}, but it is not clear how to solve the resulting coupled system of partial differential equations. However, this GKZ system is reducible which enables us to obtain its solutions individually, without having to solve the entire system at once. In the following section, we will see how reducibility allows us to do this. 

\section{Reducing GKZ systems}\label{sec:reductions}

Finding a basis of solutions for the GKZ systems introduced in the previous section is generally an involved task. Although convergent series expansions can be found algorithmically, see appendix \ref{ap:GKZ}, numerical evaluation and analytic continuation to all parameter values often present significant challenges. In this section, we will discuss how the situation can be improved when the GKZ system is reducible.
For our purposes, the reducibility of a differential system is defined as the existence of differential operators that annihilate some, but not all, of the solutions to the system. These solutions are more constrained, allowing us to find them without having to solve the entire differential system. If a sufficient number of these subsystems exist, it becomes possible to construct a complete basis of linearly independent solutions to the original differential system by solving these simpler subsystems.

We will group the additional differential operators into two classes. Firstly, there can be solutions $f$ that are independent of a coordinate $z_I$, or equivalently, satisfying $\partial_I f = 0$. From this, we obtain the first class of additional differential operators, consisting simply of the partial derivatives. Secondly, there could exist another class of operators, which we dub \textit{reduction operators} $Q_I (\nu)$, that can be of more general form. In section \ref{ssec:reductionoperators} we discuss the  subsystems obtained by using the operators $\partial_I$ and $Q_I$. In particular, we state the properties of $Q_I$ and explain how they can be obtained.  
Afterwards, in section~\ref{ssec:reducible}, we will classify when the reduction operators define non-trivial subsystems and thereby expose the precise conditions for the reducibility of a GKZ system. Note that the results of this section are obtained by describing the GKZ system as part of an Euler-Koszul homology \cite{matusevich_homological_2004}. Since these constructions can get quite technical, we will only provide the results and algorithms in this section, making it possible to use these results without having knowledge of the underlying mathematics. The reader interested in the mathematical derivations can consult appendix~\ref{ap:derivation} for details. 

\subsection{Reduction operators}\label{ssec:reductionoperators}

In this section we fix a GKZ system described by a matrix $\A$ and consider it at different parameters $\nu$. For simplicity, we assume that the number of solutions to this GKZ system, which we denote by $D$, does not depend on this parameter. Note that this assumption is valid for many examples of interest and was recently proved to hold for large classes of Feynman integrals \cite{tellander_cohenmacaulay_2023}.\footnote{Note that many of the mathematical results, such as the existence of the reduction operators, do not need this simplifying assumption. We refer to appendix~\ref{ap:derivation} for more details.}

We will now state the key properties of the reduction operators, shown in appendix~\ref{ap:derivation}. 
Let us denote a basis of linearly independent solutions to the GKZ system at parameter $\nu$ by
\beq
    f_{T}(z;\nu)\ , \qquad T=1,...,D\ . 
\eeq
In the following we assume that there are $D_I$ solutions to the GKZ system at parameter $\nu-a_I$. To indicate this dependence on the parameter we will write $D_I(\nu-a_I)$.
By assumption, these solutions satisfy 
\begin{equation}\label{eq:pift=0}
\partial_I f_t(z;\nu-a_I)=0\ , \qquad t = 1,..., D_I(\nu-a_I) \ ,
\end{equation}
where we have split the index $T=(t,\tau)$ and used $t$ as a label for those solutions annihilated by $\partial_I$. The remaining $D-D_I(\nu-a_I)$ solutions are labelled by an index $\tau$.

When assuming $D_I>0$ in~\eqref{eq:pift=0}, there are \textit{reduction operators} $Q_I (\nu)$ at $\nu$ in the direction $I$ such that:
\begin{enumerate}
    \item There are $D-D_I(\nu-a_I)$ solutions of the GKZ system at parameter $\nu$ satisfying
    \begin{equation}\label{eq:qif=0}
        Q_I (\nu)f_\tau(z;\nu)=0\ , \qquad \tau = 1,..., D-D_I(\nu-a_I)\ , 
    \end{equation}
    for all reduction operators in the direction $I$. Note that we have again labeled these solutions with $\tau$, justified by the following fact.\footnote{In general one has to be careful when considering subspaces of functions at different parameters $\nu$, as statements such as $\lim_{\nu\rightarrow \nu'} f_T(z;\nu)=f_T(z;\nu')$ are not guaranteed to hold for all $\nu$ and $\nu'$. However, we will only consider parameter shifts by the different column vectors $a_I$ which are much better behaved.}
    \item All of these solutions can be written as 
    \begin{equation}  \label{eq:relating_sol1}
    f_\tau(z;\nu)=\partial_I f_\tau(z;\nu-a_I)\ ,
    \end{equation}
    where $f_\tau(z;\nu-a_I)$ is a solution at parameter $\nu-a_I$. Note that the $f_\tau(z;\nu-a_I)$ are exactly those solutions that do not satisfy equation~\eqref{eq:pift=0}.
    \item For all solutions at $\nu$, we have 
    \begin{equation}\label{eq:relating_sol2}
        \partial_I Q_I (\nu)f_T(z;\nu)=0\, .
    \end{equation}
    Note that this is not just limited to the solutions satisfying equation~\eqref{eq:qif=0}, for which this property trivializes.
    \item The commutant of the reduction operator with the Euler operators is always of the form
        \begin{equation}\label{eq:qicommu}
        [\E_J,Q_I(\nu)]=-\left(q_I(\nu)\right)_J \,Q_I(\nu)\, 
    \end{equation}
    for some integer vector $q_I(\nu)$. If the reduction operator commutes with the toric operators~\eqref{eq:toricopdef}, this implies that there is a mapping
    \begin{equation}\label{eq:relating_sol3}
        f_T(z;\nu) \mapsto Q_I(\nu)f_T(z;\nu)
    \end{equation}
    sending solutions of the GKZ system at $\nu$ to solutions of the GKZ system at $\nu+q_I(\nu)$. Furthermore, by equation~\eqref{eq:relating_sol2}, all solutions in the image of this map will be annihilated by $\partial_I$.
\end{enumerate}
The existence of non-trivial reduction operators is determined by the value of $D_I$ and therefore depends on $\nu-a_I$. If $D_I$ is zero, the reduction operators will not reduce the solution space.
In this case, the reduction operators can still be obtained but will be proportional to the Euler operators~\eqref{eq:eulergen} and lead to trivial relations. We will now provide an algorithm for obtaining these operators if they are not trivial.

\paragraph{Obtaining reduction operators.} To derive the reduction operators at a parameter $\nu$ in the direction $I$, we have to consider the toric operators~\eqref{eq:toricopdef} determined by $\A$, and interpret them as equivalence relations. To be precise, every toric operator $\L_{u,v}$ leads to an equivalence relation of the form
\begin{equation}\label{eq:gentoricrel}
    \prod_{I=1}^N (\partial_I)^{u_I}\simeq \prod_{I=1}^N (\partial_I)^{v_I}\ .
\end{equation}
This is due to the fact that, when acting on solutions of $\L_{u,v}f=0$,
both sides of~\eqref{eq:gentoricrel} will act equivalently. When considering the set of differential operators, we introduce an equivalence relation $\simeq$ that implements~\eqref{eq:gentoricrel} by saying that two operators are equivalent if they are identical when sending $\mathcal{L}_{u,v}\rightarrow 0$. 
We will also sometimes set the partial derivative $\partial_I$ to zero, which we will denote by ${\partial_I\rightarrow 0}$. One can also combine the two and impose both $\L_{u,v}\rightarrow 0$ and $\partial_I \rightarrow 0$ and the resulting equivalence relations will be denoted by $\simeq_I$. Finally, we can generalize this to an arbitrary subset $F\subset\{1,\cdots,N\}$ and define the equivalence $\simeq_F$ which sets the partial derivatives $\partial_K\rightarrow 0$ for all $K$ in $F$, as well as the toric operators. In summary, we introduce the following notation for the equivalence relations among two differential operators $\mathcal{O}_1$, $\mathcal{O}_2$:
\begin{equation}
\begin{array}{lll}
   \mathcal{O}_1 \simeq &\hspace{-5pt} \mathcal{O}_2\  \, : \quad& (\mathcal{O}_{1}-\mathcal{O}_2)|_{\mathcal{L}_{u,v}\rightarrow 0} = 0\  ,\\
   \mathcal{O}_1 \simeq_I& \hspace{-5pt}\mathcal{O}_2\ \, : \quad &(\mathcal{O}_{1}-\mathcal{O}_2)|_{\mathcal{L}_{u,v}\rightarrow 0,\partial_I \rightarrow 0} = 0 \ ,\\
   \mathcal{O}_1 \simeq_F& \hspace{-5pt}\mathcal{O}_2\ \, :\quad &(\mathcal{O}_{1}-\mathcal{O}_2)|_{\mathcal{L}_{u,v}\rightarrow 0,\partial_K \rightarrow 0, K \in F} = 0\ . 
\end{array}
\end{equation}

The algorithm for finding reduction operators is then as follows. Start with an $M$-dimensional vector of differential operators $P_I (\nu)$, where we recall that $M$ is the number of Euler operators. We then impose that, under the equivalence relations determined by $\simeq_I$, it satisfies
\begin{equation}\label{eq:vecpi=0}
    P_I (\nu)\cdot (\E+\nu-a_I)\simeq_I 0\, ,
\end{equation}
where $\cdot$ is the vector dot product and $\E$ are the Euler operators~\eqref{eq:eulergen}. 
Note that one way a vector $P_I(\nu)$ can satisfy this equation is if its components are proportional to the Euler operators themselves. When this occurs, we refer to the resulting reduction operator as trivial, meaning it only leads to a trivial subsystem.\footnote{A subsystem is considered trivial if it either consists of the same solutions as the full system or has no non-zero solutions.} To obtain non-trivial reduction operators, one must apply the toric relations~\eqref{eq:gentoricrel}. It is always possible to find $P_I(\nu)$ that are independent of $z_I$, and we will proceed under the assumption that this has been done. Note that it is possible that $P_I(\nu)$ depends on other variables $z_J$.

Equation~\eqref{eq:vecpi=0} guarantees that, if we no longer impose that the partial derivative vanishes, we have
\begin{equation}\label{eq:vecpinot0}
\boxed{\rule[-.1cm]{0cm}{.5cm}\quad 
    P_I (\nu)\cdot (\E+\nu-a_I)\simeq Q_I (\nu)\partial_I
\quad}
\end{equation}
for some differential operator $Q_I (\nu)$. Note that the difference between equations~\eqref{eq:vecpi=0} and~\eqref{eq:vecpinot0} is that, in equation~\eqref{eq:vecpi=0}, we also impose that $\partial_I$ goes to zero. The operator $Q_I (\nu)$ obtained from equation~\eqref{eq:vecpinot0} is the reduction operator we are after. If there exists multiple operator vectors solving equation~\eqref{eq:vecpi=0}, each of them will lead to a different reduction operator.

\paragraph{Relating solutions at different parameters.} Having explained how to obtain the reduction operators, we are now ready to explain some of their properties a bit more. Note that equations~\eqref{eq:relating_sol1}, \eqref{eq:relating_sol2} and~\eqref{eq:relating_sol3} all relate the solutions of the GKZ system at one parameter to the solutions at a different parameter. This is emblematic of a general property of GKZ systems. Namely, consider a differential operator $\mathcal{O}$ that simultaneously satisfies
\begin{equation}\label{eq:comlo=0}
    [\L_{u,v},\mathcal{O}]=0\ , \qquad [\E_J,\mathcal{O}]=c_J\; \mathcal{O}\ , 
\end{equation}
for all toric operators $\L_{u,v}$, all Euler operators $\E_J$ and some  complex numbers $c_J$. For any such differential operator, one can apply it to a solution $f(z;\nu)$ of the GKZ system at $\nu$, in order to obtain a new solution at parameter $\nu-c$,
\begin{equation}
    f(z;\nu-c)\coloneqq \mathcal{O} f(z;\nu)\, ,
\end{equation}
where $c$ is the vector with components $c_J$. Of course, if $\mathcal{O}$ annihilates all solutions to the GKZ system at $\nu$, this will lead only to the trivial solutions $f(z;\nu-c)=0$.\footnote{Note that $f=0$ is always a solution to a GKZ system, since all the differential equations are homogenous.}

One example of this arises by considering a partial derivatives $\partial_I$. This clearly commutes with the toric operators while it also satisfies
\begin{equation}
    [\E_J,\partial_I]=-a_{J,I}\partial_I\ , 
\end{equation}
with $a_{J,I}$ being the element of $\A$ at the $J$-th row and the $I$-th column. This implies that it maps a solution $f$ to a solution
\begin{equation}
    f (z;\nu+a_I)\coloneqq \partial_I f(z;\nu)
\end{equation}
at $\nu+a_I$.  In fact, unless there are solutions at $\nu$ with $\partial_I f(z;\nu) =0$, all of the solutions at $\nu+a_I$ can be obtained in this way, which matches our discussion around~\eqref{eq:relating_sol1}. Crucially, it can happen that in order to reduce a GKZ system one first has to apply integer parameter shifts. The above discussion implies that these can be realized by applying partial derivatives. Or, for inverse shifts, by applying a suitable inverse operator, as described in \cite{beukers_Irreducibility_2011,bitoun_feynman_2018,caloro_ahypergeometric_2023}.

A different example is due to the fact that, as seen in equation~\eqref{eq:qicommu}, the reduction operators are guaranteed to satisfy
\begin{equation}
    [\E_J,Q_I (\nu)]=-(q_I(\nu))_J\, Q_I (\nu)
\end{equation}
for a set of integers $(q_I (\nu))_J$. This implies that, if the reduction operator also commutes with the toric operators, it can also be used to relate solutions at different parameters. In the example we consider in this paper, the reduction operators commute with the toric operators and in section~\ref{ssec:qimoving} we show how this can also be used to obtain solutions to the GKZ system. From studying various examples the property that the reduction operators commute with the toric operators seems quite common, although we have not classified exactly when this is the case.

\subsection{Existence of reductions}\label{ssec:reducible}

Reducibility of a GKZ system is not a generic property, as it is generally not expected that solutions to a set of differential equations also obey smaller, simpler subsystem.
We will now introduce the necessary and sufficient conditions for reducibility. These conditions are known in the mathematical literature \cite{walther_duality_2005,beukers_Irreducibility_2011,schulze_resonance_2012} through a property called resonance, which occurs if there are certain relations of the parameter $\nu$ with the matrix $\mathcal{A}$. In this section, we describe these results with a view towards explicit calculations. Interestingly, we will see that resonance is a geometric property and, while the exposition in this paper heavily uses the differential operators of the GKZ system, checking the resonance of a GKZ system does not require solving any differential equations.\footnote{The procedures described in this section can also be framed in a more geometric manner, here we have chosen to focus on the differential operators description instead.} Therefore, it is possible to check if reductions are possible without having to solve any subsystems explicitly. 

In general, the reducibility of a GKZ system can be checked in three steps. Leaving the precise definitions of these properties for later, these are:
\begin{enumerate}
    \item Finding a \textit{resonant set} $F \subseteq \{1,...,N \}$ for the parameter $\nu$,
    \item Determining if $F$ is a \textit{face} of $\A$,
    \item Checking that $\A$ is not a \textit{pyramid} over $F$.
\end{enumerate}
If all of these conditions are satisfied, the GKZ system is reducible and subsystems can be determined. 

\paragraph{Subsystems as GKZ systems.}  Before we give a full description of these properties, we will briefly provide some of the reasoning behind their constructions. We focus on the case where we want to find a subsystem with solutions satisfying
\begin{equation}\label{eq:pifnu=0}
    \partial_I f(z;\nu)=0 
\end{equation}
for some set of derivatives $\partial_I$. One natural thing to do is to consider the matrix $\A$ defining the original GKZ system, and pick out the column vectors $a_J$ for $J$ in some subset $F\subseteq \{1,\cdots,N\}$. If we define this matrix to be $\A_F$ we can consider the GKZ system that it defines. For example, considering $\A$ as in equation~\eqref{eq:Aexample} and $F=\{1,2,4\}$, one finds the submatrix
\begin{equation}
    \A_F= \begin{pmatrix}
    1 & 1  & 0\\
    0 & 0  & 1\\
    0 & 3  & 1\\
    1 & 2  & 0
    \end{pmatrix}
\end{equation}
of $\A$. Its solutions will be functions of the $z_J$ with $J\in F$ and therefore automatically satisfy~\eqref{eq:pifnu=0} for $I$ not in $F$. 

In fact, we show in the appendix that all solutions to the full GKZ system satisfying equation~\eqref{eq:pifnu=0} for $J\not\in F$ will also be solutions of the GKZ system defined by $\A_F$. However, the converse is not guaranteed, since the matrix $\A_F$ can have less toric operators than $\A$. This implies that it is not guaranteed that all solutions of $\A_F$ will lift to solutions of $\A$. However, if $F$ is a face of $\A$, as we will define below, then all solutions of $\A_F$ will lift to solutions of $\A$.

Another condition is also necessary to guarantee that the subsystem defined by $\A_F$ is non-trivial. Note that the matrix $\A_F$ will in general have a smaller span than the matrix $\A$. If $\nu$ is not in this span, there will be a linear combination of the Euler equations of the form $c f(z;\nu)=0$, with $c$ a non-zero constant. This implies that the combination $\A_F$ and $\nu$ will define a trivial GKZ system, with no non-zero solutions. Conversely, if
\begin{equation}
    \nu =\sum_{J \in F} c_J a_J
\end{equation}
for some complex coefficients $c_J$, the GKZ system $\A_F$ at parameter $\nu$ is non-trivial. If this is the case we say that \textit{$\nu$ is in the $\C$-span of $F$}. A generalization of this will lead to the definition of resonant sets.

Finally, it can happen that all solutions of $\A$ at parameter $\nu$ satisfy equation~\eqref{eq:pifnu=0}. If this is the case we have found a trivial subsystem which leads to no simplifications. The condition that $\A$ is not a pyramid over $F$ guarantees that this is not the case.

\paragraph{Resonant sets.} One crucial aspect of the reducibility of GKZ systems is that it is invariant under parameter shifts of the form
\begin{equation}
    \nu \rightarrow \nu+a_I\, ,
\end{equation}
where we recall that $a_I$ is a column vector of $\A$. This property is a consequence of the parameter shifts due to partial derivatives. Therefore, it is necessary to consider two parameters $\nu_1$ and $\nu_2$ equivalent if one can write
\begin{equation}
    \nu_1-\nu_2 =\sum_{I=1}^M n_I a_I\, ,
\end{equation}
where the $n_I$ are some arbitrary integers. When such $n_I$ can be found we will say that $\nu_1-\nu_2$ is in the $\Z$-span of $\A$. Often the $\Z$-span of $\A$ is the full $M$-dimensional integers $\Z^M$, in which case the equivalence relation reduces to
\begin{equation}
    \nu_1-\nu_2\in\Z^M\, .
\end{equation}
Note that this is also the case for the single-exchange system of section~\ref{ssec:singexch}.

To use this, we consider a parameter $\nu$ and a subset $F\subseteq \{1,\cdots,N\}$. We then say that $\nu$ is \textit{$F$-resonant} if there exists a parameter $\gamma$ in the $\C$-span of $\A_F$ such that $\nu-\gamma$ is in the $\Z$-span of $\A$. In other words, we require that
\begin{equation}
    \nu=\sum_{I\in F}c_I\; a_I+\sum_{J=1}^N n_J \;a_J
\end{equation}
for some combination of complex numbers $c_I$ and integers $n_J$. Alternatively, we will say that $F$ is a \textit{resonant set} for $\nu$. In practice, it is often useful to first reduce a parameter $\nu$ as much as possible modulo the $\Z$-span of $\A$, and afterwards try to combine different column vectors to check if this reduced parameter is in the $\C$-span of $F$.

\paragraph{An example of a resonant set.} Let us illustrate this with an example. Consider the GKZ system of the single exchange integral described in section~\ref{ssec:singexch}. Recall from equation~\eqref{eq:physvars} that $\nu$ takes the value of
\begin{equation}
    \nu=(\beta,\alpha)^T=(1,1,1,1+\epsilon,1+\epsilon)^T\, .
\end{equation}
Furthermore, we note that the $\Z$-span of $\A$ is $\Z^5$. Therefore, it is possible to reduce
\begin{equation}
    \nu \sim (0,0,0,1+\epsilon,1+\epsilon)^T
\end{equation}
modulo the $\Z$-span of $\A$. Now, let us now consider the subset $F=\{1,2,3,4\}$ and check if $\nu$ is $F$-resonant. Combining the column vectors for $I\in F$, it is possible to write
\begin{equation}
    (1+\epsilon)(a_2-a_1+a_4-a_3)^T=(0,0,0,1+\epsilon,1+\epsilon)^T\, .
\end{equation}
Thus we see that $(0,0,0,1+\epsilon,1+\epsilon)^T$ is in the $\C$-span of $\A_F$ and therefore $F$ is a resonant set for $\nu$.

\paragraph{Faces of $\A$.} We are now ready to discuss the second criterion necessary for obtaining subsystems. As mentioned before, this property relates to the toric operators $\L_{u,v}$ of the GKZ system, and it is encoded in their behaviour when setting $\partial_I$ to zero for each $I \not\in F$.
For each toric operator $\L_{u,v}$, one of three things can happen:
\begin{enumerate}
    \item $\L_{u,v}\vert_{\partial_I\rightarrow 0, I\not\in F}= 0$\,,
    \item $\L_{u,v}\vert_{\partial_I\rightarrow 0, I\not\in F}= \L_{u,v}$\,, 
    \item $\L_{u,v}\vert_{\partial_I\rightarrow 0, I\not\in F}=\pm \partial^w$ with $w=u$ or $w=v$\,.     
\end{enumerate}
If all toric operators are either unmodified or go to zero, i.e.~if we are in situations 1 or 2, then $F$ is a \textit{face} of $\A$. If $F$ is also a resonant set of $\nu$, we say that $F$ is a \textit{resonant face}.

For example, let us again consider the single exchange integral and $F=\{1,2,3,4\}$. The toric operators are given in equation~\eqref{eq:exchangetoric}, and inserting
\begin{equation}
    \partial_5,\, \partial_6,\, \partial_7 \rightarrow \;0
\end{equation}
we find that all toric operators go to zero. Therefore $F$ is a face and, since $F$ is $\nu$-resonant, it is also a resonant face.

As a counter example, consider the subset $G=\{2,3,4,5,7\}$. In this case, the toric operators reduce to
\begin{equation}
    \begin{array}{rl}
         \L_{u_1,v_1}|_{\partial_1 \rightarrow 0, \partial_6 \rightarrow 0} &= \partial_2 \partial_5\, ,\\
         \L_{u_2,v_2}|_{\partial_1 \rightarrow 0, \partial_6 \rightarrow 0}&= \partial_3 \partial_7 -\partial_4 \partial_5\, .
    \end{array}
\end{equation}
Since the first toric operator reduces to a monomial, $G$ is not a face of $\A$.

\paragraph{Pyramids.} Finally, we define the notion of a $\A$ being a \textit{pyramid} over a subset $F$. This is the case if all the toric operators do not involve any $\partial_I$ for $I\not \in F$. Note that if this happens, the solutions of the full GKZ system described by $\A$ are simply the solutions of $\A_F$, multiplied by a pre-factor of the form
\begin{equation}
    \prod_{I\not\in F} (z_I)^{c_I}\ ,
\end{equation}
for some complex numbers $c_I$. In particular, $\A$ and $\A_F$ have the same number of solutions and the subsystem associated to $F$ is trivial. 

We reiterate that if, for a given parameter $\nu$, it is possible to find a resonant face $F$ such that $\A$ is not a pyramid over $\A_F$, the GKZ system is reducible and it is possible to use the techniques described here and in~\ref{ssec:reductionoperators} to obtain additional differential equations satisfied by certain solutions, possibly after reducing the parameter by applying partial derivatives. With this knowledge, we are ready to try to solve reducible GKZ systems by explicitly finding the additional differential operators and solving the resulting system.

\section{Reductions of the single exchange integral}
\label{sec:singexreducs}

In this section we will apply the methods discussed above to find a full basis of solutions for the GKZ system determined in section~\ref{ssec:singexch} for the single exchange integral.
In section~\ref{ssec:reduction_operators_ex} we introduce the necessary setup for the reductions and determine the reduction operators for this system. Note that, in order to perform reductions, it is not always necessary to obtain all reduction operators. However, for this particular system, we will find that all can be obtained quite easily. In section~\ref{ssec:obtaining_solutions} we will then solve the  subsystems associated to the reduction operators and determine a full basis of solutions to the GKZ system. The latter will be used to express the result for the single exchange integral in section~\ref{ssec:explicitsols}. In section~\ref{ssec:qimoving}, we provide an alternative method for obtaining solutions, where we solve a partial derivative subsystem and then use the reduction operators to move this solution around parameter space. Finally, in the last  section~\ref{ssec:physics}, we describe the connection between the reduction operators and locality of the underlying theory.

\subsection{Determining the reduction operators}\label{ssec:reduction_operators_ex}

We begin our analysis of the single exchange GKZ system by determining the reduction operators.
To obtain the different reduction operators, recall that for each $I$, it is necessary to solve
\begin{equation}
    P_I (\nu)\cdot (\E+\nu-a_I)\simeq_I 0\, ,
\end{equation}
where $\simeq_I$ denotes that we apply the toric equivalence relations as well as set $\partial_I$ to zero. This makes it possible to obtain a reduction operator by solving
\begin{equation}\label{eq:pi=qi}
    P_I (\nu)\cdot (\E+\nu-a_I)\simeq Q_I(\nu)\partial_I
\end{equation}
for $Q_I(\nu)$, where the toric relations are now considered without $\partial_I$ set to zero. 

For convenience, let us recall that the defining data of this GKZ system is 
\beq
    \A =\left(
\begin{array}{ccccccc}
 1 & 1 & 0 & 0 & 0 & 0 & 0 \\
 0 & 0 & 1 & 1 & 0 & 0 & 0 \\
 0 & 0 & 0 & 0 & 1 & 1 & 1 \\
 0 & 1 & 0 & 0 & 0 & 1 & 0 \\
 0 & 0 & 0 & 1 & 0 & 0 & 1 \\
\end{array}
\right)\, , \qquad  \nu =\begin{pmatrix}1\\ 1\\ 1\\ 1+\epsilon\\1+\epsilon\end{pmatrix}\ . 
\eeq
Since $\A,\nu$ will be fixed throughout this section, we will omit the dependence on $\nu$ of the reduction operators and the functions, and instead write $f(z)=f(z;\nu)$, $Q_I=Q_I(\nu)$ and $P_I=P_I(\nu)$.

\paragraph{Determining $Q_1$.} Let us use this procedure to obtain $Q_1$. Recall that, for the single exchange integral, the toric operators are given by
\begin{equation}
    \begin{array}{rl}
         \L_{u_1,v_1}&=\partial_1 \partial_6-\partial_2\partial_5\\
         \L_{u_2,v_2}&=\partial_3 \partial_7 -\partial_4 \partial_5\, .
    \end{array}
\end{equation}
Upon setting $\partial_1$ to zero, these reduce to
\begin{equation}\label{eq:toricd1=0}
\begin{array}{rl}
    (\L_{u_1,v_1})|_{\partial_1\rightarrow 0} &=-\partial_2\partial_5\, ,\\
    (\L_{u_2,v_2})|_{\partial_1\rightarrow 0} &= \partial_3 \partial_7 -\partial_4 \partial_5\, .
\end{array}
\end{equation}
Furthermore, the Euler operators are also affected by the replacement $\partial_1 \rightarrow 0$. In particular, inspecting the first Euler operator in equation~\eqref{eq:exchangeeuler}, we see that
\begin{equation}
    (\E_1+(\nu-a_1)_1)|_{\partial_1\rightarrow 0}=(\theta_1+\theta_2))|_{\partial_1\rightarrow 0}= \theta_2\, ,
\end{equation}
where we recall that $\theta_I=z_I \partial_I$ 
and use $\nu-a_1=(0,1,1,1+\epsilon,1+\epsilon)^T$. Combining these observations, we find that
\begin{equation}
    \partial_5 (\E_1+ (\nu-a_1)_1) = z_1 \partial_1 \partial_5+z_2 \partial_2 \partial_5 \simeq_1 0 \ .
\end{equation}
This leads us to consider $P_1=(\partial_5,0,0,0,0)^T$ satisfying
\begin{equation}
    P_1 \cdot (\E+\nu-a_1)=z_1 \partial_1 \partial_5+z_2 \partial_2 \partial_5 \simeq_1 0\, .
\end{equation}
 $P_1$ will give rise to a reduction operator $Q_1$, which can be obtained by solving equation~\eqref{eq:pi=qi}.
 
 To obtain $Q_1$, it is necessary to apply the toric equivalence relations without setting $\partial_1 \rightarrow 0$. This yields
\begin{equation}
    P_1 \cdot (\E+\nu-a_1)\simeq (z_1 \partial_5+z_2 \partial_6)\partial_1\, ,
\end{equation}
where we have used $\partial_1 \partial_6 \simeq \partial_2 \partial_5$. From this expression, we deduce that $Q_1$ has the form
\begin{equation}
    Q_1=z_1 \partial_5+z_2 \partial_6\,.
\end{equation}
Thus we have found the reduction operator at $\nu$ in the direction $1$.

\paragraph{The complete set of reduction operators.} Repeating a similar procedure for the other values of $I$ gives rise to their associated reduction operators. In particular, choosing
\begin{equation}
    \begin{array}{ll}
        P_2 =(\partial_6,0,0,0,0)^T \, ,& P_3 =(0,\partial_5,0,0,0)^T  \, ,\\
        P_4 =(0,\partial_7,0,0,0)^T  \, ,&P_5 =(0,0,\partial_1 \partial_3,0,0)^T \, , \\
        P_6 =(0,0,\partial_2\partial_3,0,0)^T \, , &P_7 =(0,0,\partial_1\partial_4,0,0)^T \, , \\
    \end{array}
\end{equation}
results in the reduction operators
\begin{equation}\label{eq:Qidefs}
\boxed{\rule[-.1cm]{0cm}{.5cm}\quad 
\begin{array}{rl}
    Q_1=Q_2= & z_1 \partial_5+z_2 \partial_6\, , \\
    Q_3=Q_4= & z_3 \partial_5+z_4 \partial_7\, ,\\
    Q_5=Q_6=Q_7=&z_5 \partial_1 \partial_3 + z_6 \partial_2 \partial_3 +z_7 \partial_1 \partial_4\, .
\end{array}\quad}
\end{equation}

We would like to highlight how the reducibility of the GKZ system manifests itself through these reduction operators. First of all, note that each $P_I$ is zero in its last two elements. This is due to the fact that the last two elements of $\nu$ involve the arbitrary complex number $\epsilon$. One can view the operator vectors $P_I$ as determined by the relations between the different operators $\E_J+\nu_J-(a_I)_J$. Having undetermined parameters in $\nu$ then restricts which relations are possible. Secondly, many of the reduction operators are equal. This is an artefact of their associated partial derivatives not being independent, in a precise mathematical sense.\footnote{Precisely, a set of partial derivatives is considered independent if they define a regular sequence on a ring of differential operators, defined in equation~\eqref{eq:SAdef}.} As a result, these partial derivatives, and consequently their respective reduction operators, play a similar role in the GKZ system.

\subsection{Obtaining solutions using reduction operators} \label{ssec:obtaining_solutions}

To obtain the solutions associated with the different combinations of reduction operators, we first make an Ansatz similar of the form~\eqref{eq:fpgansatz}, in order to reduce the number of independent variables. Then, we can act on this Ansatz with the reduction operators in order to obtain new differential equations. The single exchange GKZ system has four linearly independent solutions and, as we will see, these are annihilated by different combinations of the reduction operators. Here, we will consider the following subsystems:
\begin{equation}\label{eq:subsystems} \boxed{\rule[-.1cm]{0cm}{1cm}\quad  
\begin{array}{lr}
     \text{subsystem $1$, satisfying:}  & Q_1f(z)=Q_3f(z)=0 \, ,\\
     \text{subsystem $2$, satisfying:}  & Q_1f(z)=Q_5f(z)=0 \, ,\\
     \text{subsystem $3$, satisfying:}  & Q_3f(z)=Q_5f(z)=0 \, ,\\
     \text{subsystem $4$, satisfying:}  & Q_5f(z)=0 \, ,
    \end{array}\quad}
\end{equation}
Note that the solutions of subsystem $2$ and $3$ will also be solutions of subsystem~$4$. However, it will be useful for us to consider these subsystems separately.

\paragraph{Solutions of subsystem $1$.} 

Denoting the solutions to the first subsystem by $f_1$, these functions must satisfy
\begin{equation}\label{eq:q1f1=0}
    Q_1f_1(z)=Q_3f_1(z)=0\, .
\end{equation}
To find these solutions, it will be useful to begin with the Ansatz
\begin{equation}\label{eq:f1ansatz}
    f_1(z)= \frac{\big(\frac{z_1 z_3}{z_2 z_4}\big)^\epsilon }{z_2 z_4 z_5}g_{1}(s,t)\, ,
\end{equation}
where $s$ and $t$ are defined in~\eqref{eq:sumvars} and $g_1$ is a yet unknown function. This Ansatz is of the type found in equation~\eqref{eq:fpgansatz}, where we have a known pre-factor combined with an unknown function of the homogeneous variables.

Inserting the Ansatz into 
\begin{equation}\label{eq:qif1=0}
    Q_1 f_1(z)=0
\end{equation}
and using the explicit expression for $Q_1$ given in equation~\eqref{eq:Qidefs}, we find that $g_{1}$ must solve
\begin{equation}
(t-1)\, \frac{\partial g_1(s,t)}{\partial t}+s \, \frac{\partial g_1(s,t)}{\partial s}+g_1(s,t)=0\, .
\end{equation}
This differential equation can be solved for $g_1$, and inserting the solution into equation~\eqref{eq:f1ansatz} results in
\begin{equation}\label{eq:Q1sol}
    f_1(z)=\frac{\big(\frac{z_1 z_3}{z_2 z_4}\big)^{\epsilon } }{z_2 z_4 z_5-z_1 z_4 z_6}\; h_1\left(\frac{z_2 z_3 z_7}{z_1 z_4 z_6-z_2 z_4 z_5}\right)\, ,
\end{equation}
where $h_1$ is yet to be determined. Note that the solutions of both the first and the second subsystem must satisfy equation~\eqref{eq:q1f1=0}. Therefore, solutions to both subsystems must be of the form~\eqref{eq:Q1sol}.

Solutions to the first subsystem will also satisfy
\begin{equation}
    Q_3  f_1(z)=0\, ,
\end{equation}
giving rise to another differential equation. Applying $Q_3$ to the solution in equation~\eqref{eq:Q1sol}, we find that $h_1$ must solve
\begin{equation}
     (u+1) \frac{\partial h_1(u)}{\partial u}+h_1(u)=0\, ,
\end{equation}
where we have defined the variable
\begin{equation}
    u\coloneqq \frac{z_2 z_3 z_7}{z_1 z_4 z_6-z_2 z_4 z_5}\, .
\end{equation}
Solving this differential equation for $h_1$ and inserting the solution into~\eqref{eq:Q1sol}, we find that the solution must be proportional
\begin{equation}\label{eq:f1sol}
 \boxed{\rule[-.1cm]{0cm}{.9cm}\quad     f_1(z)=\frac{ \big(\frac{z_1 z_3}{z_2 z_4}\big)^{\epsilon }}{z_2 z_4 z_5-z_1 z_4 z_6-z_2 z_3 z_7}
 \quad}\,.
\end{equation}
Therefore, the first subsystem has only a single linearly independent solution.

Let us higlight the simplifications that the reduction operators provided here. The full GKZ system consists of five Euler equations and two toric equations, resulting in a system of seven partial differential equations in seven variables. Applying the Ansatz of the type~\eqref{eq:generalansatz} already reduces this to only the two toric equations in two variables, however this still results in a system of two coupled second order partial differential equations which is highly nontrivial to solve directly. With the reduction operators though, it was possible to find a solution to this system by solving only two first order differential equations, both of which were easily solved using the general Ansatz. Furthermore, we note that, at any convenient point, it is possible to use the toric or Euler equations in order to solve for multiple functions at once. For example, since $f_2$ also satisfies $Q_1f_2=0$, it too takes the form of equation~\eqref{eq:Q1sol}. Inserting this into the toric equations then allows us to find both solutions at once.

\paragraph{Solving the second and third subsystem.} The process for finding the solutions to the second and third subsystem is quite similar to what we have already seen. For example, one can insert the general solution to $Q_1f_2=0$ into $Q_5f_2=0$ and find that the solutions must be proportional to
\begin{equation}\label{eq:f2sol}
 \boxed{\rule[-.1cm]{0cm}{1cm}\quad          f_{2}(z)=\frac{\left(\frac{z_1 \left(z_1 z_6-z_2 z_5\right)}{z_2^2 z_7}\right)^\epsilon}{z_2 z_4 z_5-z_1 z_4 z_6-z_2 z_3 z_7}\,.
\quad}
\end{equation}
Similarly, the equations
\begin{equation}
    Q_3  f_3(z)=Q_5  f_3(z)=0
\end{equation}
have solutions proportional to
\begin{equation}\label{eq:f3sol}
\boxed{\rule[-.1cm]{0cm}{1cm}\quad  
   f_3(z) =\frac{\left(\frac{z_3 \left(z_3 z_7-z_4 z_5\right)}{z_4^2 z_6}\right)^\epsilon}{z_2 z_4 z_5-z_1 z_4 z_6-z_2 z_3 z_7}
   \quad}\,.
\end{equation}
Note that the first three subsystems all consist of only a single linearly independent solution. Now, all that remains is to find the solutions to the last subsystem.

 \paragraph{Ans\"atze from partial solution bases.} 

Finding the solutions to the final subsystem is greatly simplified by using one of a few possible tricks available. One such trick is to use the Euler operators to rewrite the reduction operator $Q_5$ in a more convenient form. An example of this procedure is provided in section~\ref{ssec:qimoving} to obtain $f_1$, but it can be applied here just as well. However, for instructional purposes we want to highlight an additional possible simplification. Namely, that knowing some of the solutions to a set of differential equations often makes it easier to obtain the remaining solutions as well. For ordinary differential equations this technique is known as reduction of order \cite{howell_ordinary_2020} and always leads to simplifications in a very precise manner. However, in practice, one finds that such simplifications also appear often when dealing with partial differential equations.

The main idea is that, knowing a solution $f(z)$ to a differential equation, one can make an Ansatz of the form $f(z)g(z)$ and solve for $g(z)$. Since $g(z)$ being a constant must be a solution of this differential equation, the differential equation will only involve derivatives of $g(z)$, while not involving $g(z)$ itself. Defining $u(z)\coloneqq \partial g(z)/ \partial z$ this leads to a differential equation of lower order for $u$. One can then solve this and integrate the solutions to obtain the general solution. As a generalization, for a number of known solutions $f_i$, one can take an Ansatz of the form
\begin{equation}
    \sum_i f_i(z) g_i(z)
\end{equation}
where the $g_i$ now are functions to be solved for. The extra degree of freedom from the different $g_i$ can now be used to cancel other terms in the differential equation as well, providing even further simplifications. For partial differential equations these procedures can be a bit more subtle. However, even in these cases Ans\"atze such as described above still often lead to significant simplifications.

\paragraph{General solution of the fourth subsystem.}

To illustrate this, we will use the solutions $f_2$ and $f_3$ in order to obtain simpler differential equations for $f_4$. We begin by making the Ansatz
\begin{equation}
    f_4(z)=f_2(z) g_4(s,t)\, ,
\end{equation}
where $f_2$ is given in equation~\eqref{eq:f2sol} and $g_4$ is an unknown function. Inserting this Ansatz into $Q_5f_4(z)=0$ implies that $g_4$ must solve
\begin{equation}
   \left( \epsilon\,(2 s-1) +(s-1) s\, \frac{\partial}{\partial s}\right) \frac{\partial g_4(s,t)}{\partial t}=0\, .
\end{equation}
Note that defining $u(s,t)\coloneqq\partial g_4(s,t)/\partial t$ this becomes an ordinary differential equation for $u$. Solving this equation results in
\begin{equation}
    g(s,t)=h_1(s)+\frac{h_2(t)}{(s-s^2)^{\epsilon}}\, ,
\end{equation}
where $h_1$ and $h_2$ are undetermined functions. Inserting this into into the expression for $f_4$ and performing some function redefinitions for $h_1$ and $h_2$, this implies that the general solution to $Q_5f_4(z)=0$ can be written as
\begin{equation}\label{eq:f4h12ansatz}
    f_4(z)=f_2(z) h_1(s)+f_3(z)h_2(t)\, ,
\end{equation}
where $f_3$ is as in equation~\eqref{eq:f3sol}.

\paragraph{Obtaining the remaining linearly independent solution.} 
Using the general solution to $Q_5f_4(z)=0$, we will now obtain the fourth and final basis function, providing us with a full basis of linearly independent solutions to the GKZ system. The advantage of the Ansatz~\eqref{eq:f4h12ansatz} is that the resulting differential equations only involve the derivatives of $h_1$ and $h_2$, while not involving $h_1$ and $h_2$ directly. For example, the first toric equation
\begin{equation}
    \L_{u_1,v_1}f_4(z)=0
\end{equation}
reduces to a differential equation of the form
\begin{equation}\label{eq:h1h2eq}
    p_1(s,t) \frac{\partial h_1(s)}{\partial s}+p_2(s,t) \frac{\partial^2 h_1(s)}{\partial s^2}+p_3(s,t) \frac{\partial h_2(t)}{\partial t}=0\, ,
\end{equation}
where the $p_i$ are polynomials in $s,t$ and $\epsilon$. Solving for $\partial h_2/\partial t$ and imposing that
\begin{equation}
    \frac{\partial^2 h_2(t)}{\partial s\partial t }=0
\end{equation}
results in a third order differential equation for $h_1(t)$. Or, since the differential equation does not involve $h_1(t)$ but only its derivatives, a second order differential equation for $\partial h_1(t)/\partial t$. The solution to this differential equation is given by
\begin{equation}
    h_1(t)=\frac{c_1\,_2F_1(1,-2\epsilon,1-\epsilon;1-s)}{(s-s^2)^{\epsilon } }+\frac{c_2 }{\left(s^2-s\right)^{\epsilon }}+c_3\, ,
\end{equation}
where the $c_i$ are undetermined constants and $_2F_1$ is the hypergeometric function. However, inserting this into expression~\eqref{eq:f4h12ansatz} for $f_4$ one finds that the term proportional to $c_2$ can be absorbed into $h_1(s)$, while $c_3$ reproduces a solution proportional to $f_2$. Since we are only interested in finding new solutions to the GKZ system, we will set $c_2=c_3=0$ from now on.

Inserting the solution for $h_2$ into equation~\eqref{eq:h1h2eq} results in the first order inhomogenous differential equation
\begin{equation}
    \frac{\partial h_2(t)}{\partial t}=\frac{-\epsilon \,c_1}{ (t-t^2)^{\epsilon}}
\end{equation}
for $h_2$, with as its solution
\begin{equation}
    h_2(t)=\frac{c_1\,_2F_1(1,-2\epsilon,1-\epsilon;1-t)}{(t-t^2)^{\epsilon } }+c_4\, ,
\end{equation}
where $c_4$ is a new integration constant. Again, inserting this in to our expression for $f_4$ one finds that the $c_4$ term is proportional to $f_3$ and therefore we will set $c_4=0$ as well. Finally, we will insert the expressions for $h_1$ and $h_2$ into equation~\eqref{eq:f4h12ansatz} resulting in
\begin{equation}\label{eq:f4sol}
\boxed{\rule[-.1cm]{0cm}{1cm}\quad  
\begin{aligned}
    f_4(z)=& \quad\frac{\left(\frac{z_5^2}{z_6 z_7}\right){}^\epsilon \, _2F_1\left(1,-2 \epsilon ;1-\epsilon ;1-\frac{z_1 z_6}{z_2 z_5}\right)}{z_1 z_4 z_6+z_2 z_3 z_7-z_2 z_4 z_5}\\
    &+\frac{\left(\frac{z_5^2}{z_6 z_7}\right){}^\epsilon \, _2F_1\left(1,-2 \epsilon ;1-\epsilon ;1-\frac{z_3 z_7}{z_4 z_5}\right)}{z_1 z_4 z_6+z_2 z_3 z_7-z_2 z_4 z_5}\, ,
\end{aligned}\quad}
\end{equation}
where, since we are only interested in linearly independent solutions, we have set $c_1=1$.

With the four functions~\eqref{eq:f1sol},~\eqref{eq:f2sol},~\eqref{eq:f3sol}, and~\eqref{eq:f4sol} we have a full basis of solutions to the GKZ system at the parameter $\nu$ and, since the original integral in equation~\eqref{eq:singleexchange} is a solution to this system of differential equations, it should be some linear combination of these four functions. We will now provide these coefficients, as well as the explicit form of the solution in terms of the physical variables $X_1$, $X_2$ and $Y$.

\subsection{Solutions in terms of the physical variables}\label{ssec:explicitsols}

Having found the four different functions spanning the solution space of our GKZ system in section~\ref{ssec:obtaining_solutions}, it is now possible to write any convergent integral of the type ~\eqref{eq:exchangegkz} with $\nu=(1,1,1,1+\epsilon,1+\epsilon)^T$ as
\begin{equation}\label{eq:igammagensol}
    I_\Gamma(z;\nu)=c_1(\Gamma;\epsilon)f_1(z)+c_2(\Gamma;\epsilon)f_4(z)c_3(\Gamma;\epsilon)f_{\nu,3}(z)+c_4(\Gamma;\epsilon)f_4(z)\, ,
\end{equation}
where the functions $f_{i}$ are given in equations~\eqref{eq:f1sol},~\eqref{eq:f2sol},~\eqref{eq:f3sol} and~\eqref{eq:f4sol} while the coefficients $c_i(\Gamma;\epsilon)$ can be obtained fixing $\epsilon$ and an integration cycle $\Gamma$, or analytically by considering the integral in specific limits and considering the integral in certain limits. 

The single exchange integral~\eqref{eq:singleexchange} is a special case of the function~\eqref{eq:igammagensol}, obtained by setting the variables $z_I$ to their physical values from equation~\eqref{eq:physvars} and taking the integration cycle $\R^2_+$. With these replacements, the basis functions take the form
\begin{equation}\label{eq:gensols}
\begin{split}
    f_1(z)\vert_{\mathrm{phys}} &=\frac{(X_1+Y)^{\epsilon } (X_2+Y)^{\epsilon }}{2 Y} \, ,\\
     f_2(z)\vert_{\mathrm{phys}} &=\frac{(X_1+Y)^{\epsilon } (X_2-Y)^{\epsilon }}{2 Y}\, , \\
     f_3(z)\vert_{\mathrm{phys}} &=\frac{(X_1-Y)^{\epsilon } (X_2+Y)^{\epsilon }}{2 Y}\, ,\\
     f_4(z)\vert_{\mathrm{phys}} &= \frac{(X_1+X_2)^{2 \epsilon } \left(\, _2F_1\left(1,-2 \epsilon ;1-\epsilon ;\frac{X_1-Y}{X_1+X_2}\right)\right)}{2 Y}\\
     &+ \frac{(X_1+X_2)^{2 \epsilon } \left(\, _2F_1\left(1,-2 \epsilon ;1-\epsilon ;\frac{X_2-Y}{X_1+X_2}\right)\right)}{2 Y}\, ,\\
\end{split}
\end{equation}
where $\vert_\mathrm{phys}$ means that we replace the $z_I$ with their physical values. 

The different coefficients can be obtained by evaluating the functions above, as well as the single exchange integral in certain limits of $X_1$, $X_2$ and $Y$. This results in
\begin{equation}\label{eq:coefssols}
\begin{array}{rl}
     c_1(\R_+^2;\epsilon) &= -\pi ^2 \csc ^2(\pi  \epsilon )\, ,\\
     c_2(\R_+^2;\epsilon) &=0\, ,\\
     c_3(\R_+^2;\epsilon) &=0\, ,\\
     c_4(\R_+^2;\epsilon) &= 2^{-2 \epsilon -1} \sqrt{\pi } \csc (\pi  \epsilon ) \Gamma \left(\frac{1}{2}-\epsilon \right) \Gamma (\epsilon )\, ,
\end{array}
\end{equation}
for the different coefficients. Inserting all of the above into equation~\eqref{eq:igammagensol} we obtain an expression for the single exchange integral. This expression is in agreement with \cite{arkani-hamed_differential_2023} after applying a series of hypergeometric identities. Interestingly, the reduction operators also imply a series of inhomogeneous equations satisfied by the single-exchange integral. In turn, these also lead to boundary conditions using which the coefficients $c_i$ can be obtained as well. We will discuss this in more detail in section~\ref{ssec:physics}.

\subsection{An alternative: relating subsystems using reduction operators}\label{ssec:qimoving}

Before we go on discussing the relation between the reduction operators and locality, we want to highlight an alternative method that one can use to obtain solutions using the reduction operators. This method only works if the reduction operator commutes with the toric operators. However, one finds that this is the case in many examples, including our main example of section \ref{ssec:singexch}. 

Recall from section~\ref{ssec:reductionoperators} that, if a reduction operator commutes with the toric operators, it provides a map 
\begin{equation}\label{eq:qimap}
    f(z;\nu)\mapsto Q_If(z;\nu)
\end{equation}
between solutions of the GKZ system at $\nu$ to solutions at some different parameter. And this parameter can be obtained by considering the commutation relations
\begin{equation}
    [\E_J,Q_I(\nu)]=-(q_I(\nu))_J Q_I(\nu)\, ,
\end{equation}
where $q_I$ is an integer vector. 

One reason this map is useful is that the reduction operators satisfy
\begin{equation}
    \partial_I Q_I(\nu)f(z;\nu)=0\, ,
\end{equation}
for all solutions $f$ of the GKZ system at $\nu$. This implies that the image of the map~\eqref{eq:qimap} consists of solutions $\tilde f$ of the GKZ system at $\nu+q_I(\nu)$ satisfying
\begin{equation}
    \partial_I \tilde f(z;\nu+q_I(\nu))=0\, .
\end{equation}
In the notation introduced at the beginning of section \ref{ssec:reductionoperators}, the $\tilde f$ is among the solutions with index $t$. These relations provide the link between the two different kinds of subsystems. 
Therefore, if we can find a solution $\tilde f(z;\nu+q_I(\nu))$ at $\nu+q_I(\nu)$ that is independent of $z_I$, it is possible to obtain new solutions $f$ at $\nu$ by solving the inhomogenous differential equation
\begin{equation}\label{eq:qif=ft}
    Q_I(\nu)f(z;\nu)=\tilde f(z;\nu+q_I(\nu))\ .
\end{equation}

\paragraph{Solving partial derivative subsystem for $f_1$.}

To put this into practice, let us consider the function $f_1$ from equation~\eqref{eq:f1sol}. Inspecting the different subsystems in equation~\eqref{eq:subsystems}, we find that this is the only solution of the GKZ system satisfying
\begin{equation}
    Q_5f(z)\neq 0\, .
\end{equation}
From the discussion above, we see that this implies that
\begin{equation}\label{eq:q6f1=ft}
    Q_5f_1(z)=\tilde f(z;\nu+q_5)\ ,
\end{equation}
for some solution $\tilde f$ of the GKZ system at $\nu+q_5$. Furthermore, we have that $\tilde f$ must satisfy
\begin{equation}\label{eq:d6ft=0}
    \partial_6 \tilde f(z;\nu+q_5)=0
\end{equation}
by the properties of the reduction operators. Finally, since $Q_5$ satisfies
\begin{equation}
    [\E_J,Q_5]=-(a_1+a_3-a_5)Q_5
\end{equation}
with $a_I$ the column vectors of $\A$, we find that
\begin{equation}
    q_5=a_1+a_3-a_5\,.
\end{equation}
This implies that we should consider solutions of the GKZ system at the parameter
\begin{equation}
    \tilde{\nu}\coloneqq \nu+q_5=(2,2,0,1+\epsilon,1+\epsilon)^T\,.
\end{equation}
In particular, note that $\tilde{\nu}_3=0$.

It is now possible to try to solve for $\tilde f$ and insert it into equation~\eqref{eq:q6f1=ft} to obtain $f_1$. However, there is some further simplification we can do. Since the reduction operators satisfy
\begin{equation}
    Q_5=Q_6=Q_7\, ,
\end{equation}
we have that
\begin{equation}
     Q_5f_1(z)=Q_6 f_1(z)=Q_7f_1(z)=\tilde f(z;\tilde{\nu})\,.
\end{equation}
By the same arguments as before, this implies that $\tilde f$ must not only satisfy equation~\eqref{eq:d6ft=0}, but also
\begin{equation}
    \partial_6 \tilde f(z;\tilde{\nu})=\partial_7 \tilde f(z;\tilde{\nu})=0
\end{equation}
constraining $\tilde f$ further.

Recall from section~\ref{ssec:reducible} that, when studying solutions of the GKZ system annihilated by partial derivatives, it is possible to interpret these as solutions to a smaller GKZ system. For a set of partial derivatives $\{I_1,\cdots,I_k\}$, this smaller system is obtained by defining the matrix $\A_F$ from the columns $a_I$ of $\A$. Where we take only the columns
\begin{equation}
    I \in \{1,\cdots,N\}\setminus \{I_1,\cdots,I_k\}\,,
\end{equation}
with $N$ being the number of columns of $\A$. The solutions of the original GKZ system annihilated by these partial derivatives will then also be solutions of the GKZ system defined by $\A_F$.

To obtain $\tilde f$, we must therefore construct the matrix with columns
\begin{equation}
    \{1,2,3,4,5,6,7\}\setminus \{5,6,7\}=\{1,2,3,4\}\,.
\end{equation}
In other words, the matrix $\A_F$ consisting of only the first four columns of $\A$. This matrix is given by
\begin{equation}\label{eq:Afmat}
    \A_{F}=\begin{pmatrix}
        1 & 1& 0&0\\
        0&0 &1&1\\
        0&0&0&0\\
        0&1&0&0\\
        0&0&0&1
    \end{pmatrix}\, ,
\end{equation}
and we will now consider the GKZ system it defines. It turns out that this GKZ system is especially simple as it has no toric operators, while the Euler operators are
\begin{equation}
        \begin{array}{lll}
       \tilde \E_1= \theta_1+\theta_2\, , & \tilde \E_2=\theta_3 +\theta_4\, , & \tilde\E_3=0\ ,\\
        \tilde\E_4=\theta_2 \, , & \tilde\E_5=\theta_4\, . &
    \end{array}
\end{equation}
Therefore, $\tilde{f}$ is a solution to the system
\beq\label{EulerF-system}
   (\tilde \E_J + \tilde \nu_J)\tilde f(z;\tilde{\nu}) = 0\ ,
\eeq
with $1\leq J\leq 5$. Note that since $\tilde \E_3=0$ the above implies  $\tilde{\nu}_3\tilde f(z;\tilde{\nu})=0$. Thus, if $ \tilde{\nu}_3\neq 0$,
the system will have no non-zero solutions and the GKZ system is trivial. This exemplifies why a parameter must be in the $\C$-span of $F$ to obtain non-trivial subsystems.

The solution to the differential equations \eqref{EulerF-system} can be found quite easily, and its one-dimensional solution space is spanned by the function 
\begin{equation}
    \tilde{f}(z;\tilde{\nu})=\frac{\big(\frac{z_1 z_3}{z_2 z_4}\big)^{\epsilon}}{z_1 z_2 z_3 z_4}\, .
\end{equation}
Therefore, to obtain $f_1(z)$, we must solve
\begin{equation}\label{eq:q6f=zz}
    Q_5f_1(z)=\frac{\big(\frac{z_1 z_3}{z_2 z_4}\big)^{\epsilon}}{z_1 z_2 z_3 z_4}\, .
\end{equation}
Before we do this, we will showcase a useful way that reduction operators can be rewritten, which greatly simplifies solving equation \eqref{eq:q6f=zz}.

\paragraph{Rewriting reduction operators.} 

As it stands, equation \eqref{eq:q6f=zz} is a bit inconvenient, as it is somewhat involved to solve this differential equation directly. However, we will use this to showcase another useful tool that can be used to simplify such differential equations. The main idea is that, like before, we will consider various differential operators to be equivalent if they act on solutions of the GKZ system in equivalent ways. We have already used this for the toric equations, and the difference here is that we will also apply this reasoning to the Euler equations. 

Consider the Euler equations
\begin{equation}\label{eq:eulerrep}
    \begin{split}
        \E_1+\nu_1&=\theta_1+\theta_2+1\, ,\\
        \E_2+\nu_2&=\theta_3+\theta_4+1\, ,
    \end{split}
\end{equation}
where we recall that $\theta_i=z_i\partial_i$. Note that, by commuting $z_I$ and $\partial_I$, one finds that
\begin{equation}
    \theta_I+1=\partial_I z_I\, ,
\end{equation}
where the derivative acts on everything to its right. Combining this observation with equation~\eqref{eq:eulerrep} and solving for $\partial_4$ and $\partial_2$, one finds that
\begin{equation}\label{eq:dieq}
\begin{array}{rl}
   \partial_4 &\simeq_{\E+\nu} -\partial_3 \frac{z_3}{z_4}\, ,\\
   \partial_2 & \simeq_{\E+\nu} -\partial_1 \frac{z_1}{z_2}\, ,
\end{array}
\end{equation}
where, $\simeq_{\E+\nu}$ means that we consider equivalence relations stemming from the toric equations, as well as the Euler equations. In other words, both sides of equation~\eqref{eq:dieq} act the same on solutions of the GKZ system.

With these replacements, it is possible rewrite the reduction operator $Q_5 (\nu)$ as\footnote{These kind of replacements can also be useful for finding representations of reduction operators that commute with the toric operators, since this property can depend on the representation.}
\begin{equation}\label{eq:tildeq5def}
    Q_5  \simeq_{\E+\nu}  \frac{1}{z_2 z_4}   \partial_1 \partial_3 (z_2 z_4 z_5-z_1 z_4 z_6-z_2 z_3 z_7) \eqqcolon \tilde{Q}_6\, .
\end{equation}
As we will see, this representation simplifies the process of solving for $f_1$ significantly.

\paragraph{Inhomogenous equations from reduction operators.}

Recall that we are interested in solving equation~\eqref{eq:q6f=zz} for $f_1(z)$. By the discussion above, it is possible to replace $Q_5$ with $\tilde{Q}_6$ when acting on $f_1$, leading to the differential equation
\begin{equation}
 \frac{1}{z_2 z_4}   \partial_1 \partial_3 \big((z_2 z_4 z_5-z_1 z_4 z_6-z_2 z_3 z_7) f_1(z) \big)=\frac{\left(\frac{z_1 z_3}{z_2 z_4}\right)^{\epsilon}}{z_1 z_2 z_3 z_4}\, .
\end{equation}
Since the left hand side consists of total derivatives with respect to $\partial_1$ and $\partial_3$, this equation can simply be integrated twice resulting in
\begin{equation}
        f_1(z)=\frac{ \left(\frac{z_1 z_3}{z_2 z_4}\right){}^{\epsilon }}{z_2 z_4 z_5-z_1 z_4 z_6-z_2 z_3 z_7}
\end{equation}
consistent with equation~\eqref{eq:f1sol}. Note that in this integration two integration constants have been ignored. Taking these into account properly would result in including the solutions $Q_5 f=0$ as well. However, since we are only interested in a particular solution, we can simply ignore these. 

Especially when convenient representations of differential operators can be found, solving inhomogenous differential equations of the type~\eqref{eq:qif=ft} is a way one can easily obtain solutions to the GKZ system. Alternatively, if one reduction operator is much simpler than the others, or if only one exists, this technique makes it possible to only consider differential equations incorporating this single reduction operator. Since solving the inhomogenous differential equation in full generality includes all functions annihilated by the reduction operator, as well as those that are mapped to the inhomogenous part.

\subsection{Locality as a consequence of reduction operators} \label{ssec:physics}

As we saw in section~\ref{sec:ReviewCC}, the single exchange integral satisfies a set of differential equations in $X_1$ and $X_2$ separately, due to the locality of the underlying theory. In this section we show that this property is a direct consequence of the existence of the reduction operators $Q_1 $ and $Q_3 $. The framework of reduction operators also allows us to classify when these kind of equations can exist depending on which components of the parameter $\nu$ are non-integer, providing an alternative perspective to a similar result obtained in \cite{arkani-hamed_differential_2023}. We then speculate about some how this can be generalized. To do this, we first discuss how acting with reduction operators can simplify an integral.

\paragraph{Reduced integrals from reduction operators.}

Recall from section~\ref{ssec:qimoving} that the acting of the reduction operators on a solution maps it to the solution of a smaller GKZ system. Since solutions to GKZ systems can often be described by integrals, it is natural to try to construct the integrals associated to this smaller GKZ system. Note that it is not guaranteed that such integrals provide us with a complete basis of solutions. However, if the parameter is non-resonant for the matrix defining the smaller GKZ system, this is guaranteed.\footnote{Note that a parameter $\nu$ can be simultaneously resonant for $\A$ as well as non-resonant for a submatrix $\A_F$.} 

As an example, let us consider the subsystem considered in section~\ref{ssec:qimoving}, defined from the matrix~\eqref{eq:Afmat}. To construct the associated integrals, one can simply reverse the process described in section~\ref{ssec:gengkz} and identify the polynomials
\begin{equation}
    p_1=z_1+z_2 \omega_1\, ,\qquad p_2=z_3+z_4 \omega_2\,,
\end{equation}
which give rise to the matrix $\A_F$. Note that rows filled with zeroes play no role in this construction. From these polynomials we obtain integrals of the form
\begin{equation}
    \int_\Gamma d^2\omega\frac{(\omega_1\omega_2)^\epsilon}{(z_1 +z_2 \omega_1)^2(z_3 +z_4 \omega_2)^2}
\end{equation}
for the associated GKZ system.

\paragraph{Reduced integrals for the single-exchange integral.} One can apply the same reasoning to the other reduction operators $Q_1 $ and $Q_3 $. With this, one finds the following identities for the single exchange integral:
\begin{subequations}\label{eq:Qactions}
    \begin{align}
        Q_1  I_{\R_+^2}(z;\nu) \vert_{\rm phys}&=-\epsilon \int_{\R_+^2}d\omega_1 d\omega_2\; \frac{(\omega_1 \omega_2)^\epsilon}{(\omega_2+X_2+Y)(\omega_1+\omega_2+X_1+X_2)^2}\, , \label{eq:Q1action}\\
        Q_3  I_{\R_+^2}(z;\nu) \vert_{\rm phys}&=-\epsilon \int_{\R_+^2}d\omega_1 d\omega_2\; \frac{(\omega_1 \omega_2)^\epsilon}{(\omega_1+X_1+Y) (\omega_1+\omega_2+X_1+X_2)^2}\, ,  \label{eq:Q3action}\\
        Q_5  I_{\R_+^2}(z;\nu) \vert_{\rm  phys}&=\epsilon^2 \int_{\R_+^2}d\omega_1 d\omega_2\; \frac{(\omega_1 \omega_2)^\epsilon}{(\omega_1+X_1+Y)^2 (X_2+\omega_2+Y)^2}\, ,  \label{eq:Q5action}\\
        Q_3  Q_1  I_{\R_+^2}(z;\nu) \vert_{\rm phys}&=\epsilon^2 \int_{\R_+^2}d\omega_1 d\omega_2\; \frac{(\omega_1 \omega_2)^\epsilon}{ (\omega_1+\omega_2+X_1+X_2)^3}\, ,  \label{eq:Q13action}
    \end{align} 
\end{subequations}
where $I_{\R_+^2}^2(z;\nu)$ is as defined in equation~\eqref{eq:exchangegkz} and we have specialized the coordinates $z_I$ to their physical values after applying the differential operators. The pre-factors of $-\epsilon$ and $\epsilon^2$ do not follow immediately from this reasoning but have instead been obtained by explicit calculation. It turns out that the right-hand sides of these equations have some interesting physical interpretations.

\begin{figure}[ht]
\centering
    \begin{subfigure}{0.49\textwidth}
    \centering
    \begin{tikzpicture}
        \begin{feynman}
            \vertex (topleft);
            \vertex [right=.75 cm of topleft,boundarydot] (topcenter1) {};
            \vertex [right=1.5 cm of topcenter1,boundarydot] (topcenter2) {};
            \vertex [right=.75 cm of topcenter2,boundarydot] (topcenter);
            \vertex [below=1.5 cm of topcenter] (midcenter) ;
            \vertex[left=1.5 cm of midcenter,bulkdot] (leftdiagram) {};
            \vertex[right=1.5 cm of midcenter,bulkdot] (rightdiagram){};
            \vertex [right=.75 cm of topcenter,boundarydot] (topcenter3) {};
            \vertex [right=1.5 cm of topcenter3,boundarydot] (topcenter4) {};
            \vertex [right =.75 cm of topcenter4] (topright);
    
            \diagram*  {
              (topleft) --[very thick,blue] (topcenter1) --[very thick,blue] (topcenter2) --[very thick,blue] (topcenter3) --[very thick,blue] (topcenter4) --[very thick,blue] (topright);
              (leftdiagram) --[thick,opacity=1] (topcenter1);
              (leftdiagram) --[thick,opacity=1]  (topcenter2);
              (rightdiagram) --[thick,opacity=1] (topcenter3);
              (rightdiagram) --[thick,opacity=1] (topcenter4);
              (rightdiagram) --[dashed,very thick,green,with arrow=.5] (leftdiagram);
            };
            \vertex[below=.5 cm of leftdiagram] (X1) {\textbf{$X_1$}};
            \vertex[below=.5 cm of rightdiagram] (X2) {\textbf{$X_2$}};
            \vertex[above=0cm of midcenter] (Y) {\textbf{\color{green} $Y$}};
        \end{feynman}
    \end{tikzpicture}
    \caption{}
    \vspace*{5mm}
    \end{subfigure}
    \hfill
    \begin{subfigure}{0.49\textwidth}
    \centering
    \begin{tikzpicture}
        \begin{feynman}
            \vertex (topleft);
            \vertex [right=.75 cm of topleft,boundarydot] (topcenter1) {};
            \vertex [right=1.5 cm of topcenter1,boundarydot] (topcenter2) {};
            \vertex [right=.75 cm of topcenter2,boundarydot] (topcenter);
            \vertex [below=1.5 cm of topcenter] (midcenter) ;
            \vertex[left=1.5 cm of midcenter,bulkdot] (leftdiagram) {};
            \vertex[right=1.5 cm of midcenter,bulkdot] (rightdiagram){};
            \vertex [right=.75 cm of topcenter,boundarydot] (topcenter3) {};
            \vertex [right=1.5 cm of topcenter3,boundarydot] (topcenter4) {};
            \vertex [right =.75 cm of topcenter4] (topright);
    
            \diagram*  {
              (topleft) --[very thick,blue] (topcenter1) --[very thick,blue] (topcenter2) --[very thick,blue] (topcenter3) --[very thick,blue] (topcenter4) --[very thick,blue] (topright);
              (leftdiagram) --[thick,opacity=1] (topcenter1);
              (leftdiagram) --[thick,opacity=1]  (topcenter2);
              (rightdiagram) --[thick,opacity=1] (topcenter3);
              (rightdiagram) --[thick,opacity=1] (topcenter4);
              (leftdiagram) --[dashed,very thick,green,with arrow=.5] (rightdiagram);;
            };
            \vertex[below=.5 cm of leftdiagram] (X1) {\textbf{$X_1$}};
            \vertex[below=.5 cm of rightdiagram] (X2) {\textbf{$X_2$}};
            \vertex[above=0cm of midcenter] (Y) {\textbf{\color{green} $Y$}};
        \end{feynman}
    \end{tikzpicture}
    \caption{}
    \vspace*{5mm}
    \end{subfigure}
    
    \begin{subfigure}{0.49\textwidth}
    \centering
    \begin{tikzpicture}
        \begin{feynman}
            \vertex (topleft);
            \vertex [right=.75 cm of topleft,boundarydot] (topcenter1) {};
            \vertex [right=1.5 cm of topcenter1,boundarydot] (topcenter2) {};
            \vertex [right=.75 cm of topcenter2,boundarydot] (topcenter);
            \vertex [below=1.5 cm of topcenter] (midcenter) ;
            \vertex[left=1.5 cm of midcenter,bulkdot] (leftdiagram) {};
            \vertex[right=1.5 cm of midcenter,bulkdot] (rightdiagram){};
            \vertex [right=.75 cm of topcenter,boundarydot] (topcenter3) {};
            \vertex [right=1.5 cm of topcenter3,boundarydot] (topcenter4) {};
            \vertex [right =.75 cm of topcenter4] (topright);
    
            \diagram*  {
              (topleft) --[very thick,blue] (topcenter1) --[very thick,blue] (topcenter2) --[very thick,blue] (topcenter3) --[very thick,blue] (topcenter4) --[very thick,blue] (topright);
              (leftdiagram) --[thick,opacity=1] (topcenter1);
              (leftdiagram) --[thick,opacity=1]  (topcenter2);
              (rightdiagram) --[thick,opacity=1] (topcenter3);
              (rightdiagram) --[thick,opacity=1] (topcenter4);
            };
            \vertex[below=.5 cm of leftdiagram] (X1) {\textbf{$X_1 +$\textbf{\color{green} $Y$}}};
            \vertex[below=.5 cm of rightdiagram] (X2) {\textbf{$X_2 +$\textbf{\color{green} $Y$}}};
        \end{feynman}
    \end{tikzpicture}
    \caption{}\label{eq:singleexchangecut}
    \end{subfigure}
    \hfill
    \begin{subfigure}{0.49\textwidth}
    \centering
    \begin{tikzpicture}
        \begin{feynman}
            \vertex (topleft);
            \vertex [right=.75 cm of topleft,boundarydot] (topcenter1) {};
            \vertex [right=1.5 cm of topcenter1,boundarydot] (topcenter2) {};
            \vertex [right=.75 cm of topcenter2,boundarydot] (topcenter);
            \vertex [below=1.44 cm of topcenter,bulkdot] (midcenter) {};
            \vertex [right=.75 cm of topcenter,boundarydot] (topcenter3) {};
            \vertex [right=1.5 cm of topcenter3,boundarydot] (topcenter4) {};
            \vertex [right =.75 cm of topcenter4] (topright);
    
            \diagram*  {
              (topleft) --[very thick,blue] (topcenter1) --[very thick,blue] (topcenter2) --[very thick,blue] (topcenter3) --[very thick,blue] (topcenter4) --[very thick,blue] (topright);
              (midcenter) --[thick,opacity=1] (topcenter1);
              (midcenter) --[thick,opacity=1]  (topcenter2);
              (midcenter) --[thick,opacity=1] (topcenter3);
              (midcenter) --[thick,opacity=1] (topcenter4);;
            };
            \vertex[below=0.5cm of midcenter] (Xs) {\textbf{$X_1+X_2$}};
        \end{feynman}
    \end{tikzpicture}
    \caption{}
    \end{subfigure}
\caption{The diagrammatical interpretation of the integrals in equation~\eqref{eq:Qactions}. In particular, we have that (a) corresponds to the left-ordered part of the propagator, (b) to the right-ordered part and (c) to the non-time-ordered part. Finally, (d) corresponds to the collapsed propagator.} \label{fig:Qdiagrams}
\end{figure}
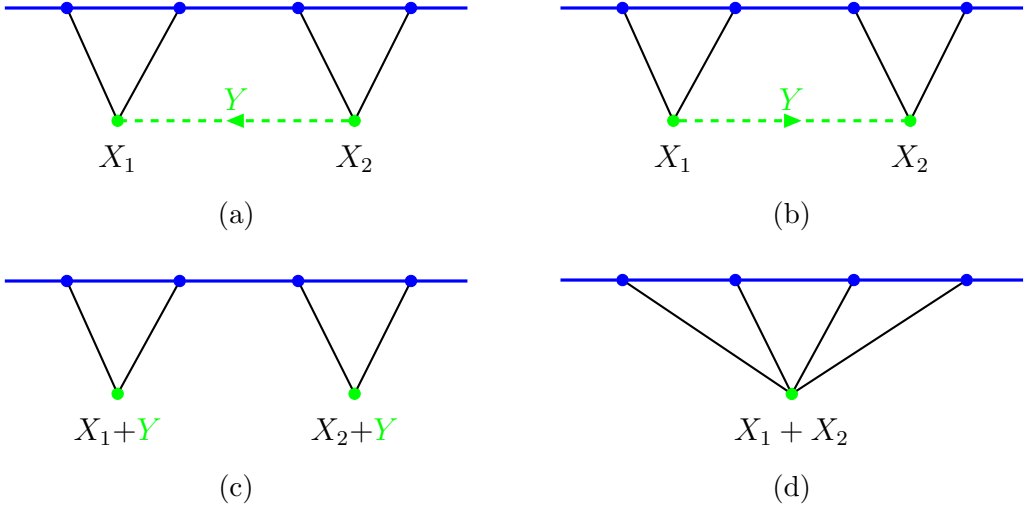

The different integrals above can be interpreted diagrammatically, due to the fact that the propagator $G(Y,\eta_1,\eta_2)$ in equation~\eqref{eq:prop-explicit} has three terms. The first two terms both come with Heaviside step function enforcing either $\eta_1\leq \eta_2$ or $\eta_1 \geq \eta_2$, we will call these the left-ordered or right-ordered parts respectively. The third term does not come with a Heaviside step function and therefore we will call it the non-ordered part. This lack of time ordering implies that the two vertices disconnect and the resulting integral is just a product of the two vertices. Interestingly, if one considers these terms separately and performs the integrations over $\eta_1$ and $\eta_2$, one recovers the right-hand sides of equations~\eqref{eq:Q1action},~\eqref{eq:Q3action} and~\eqref{eq:Q5action} respectively, up to a twist in the powers of the different polynomials. This twist can be realized by acting with derivatives of $X_1$ or $X_2$. The integral in equation~\eqref{eq:Q13action} also has a diagrammatic interpretation. This integral simply corresponds to collapsing the propagator to a point, leading to a single vertex as shown in (d) of figure~\ref{fig:Qdiagrams}.

\paragraph{Locality from reduction operators.} The collapsed propagator is particularly interesting since this also appeared in the discussion of locality in section~\ref{sec:ReviewCC}. Here, collapsing the propagator implied a second order inhomogeneous differential equation for the single exchange integral. Furthermore this differential equation was quite special as it depended on only one of $X_1$ or $X_2$. As we will see, equation~\eqref{eq:Q13action} is exactly equation~\eqref{eq:locality}, only rewritten in terms of $Q_1 $ and $Q_3 $ instead of the derivatives of $X_1$. In particular, the right hand side of equation~\eqref{eq:locality} is exactly the integral appearing in~\eqref{eq:Q13action}, where the twists are realized by the partial derivatives of $X_1$. Therefore it is possible to rewrite
\begin{equation}\label{eq:Q13nu=ict}
    Q_3 Q_1 I_\epsilon \vert_{\mathrm{phys}}=\frac{\partial^2 I_{\mathrm{contr}}}{\partial X_1^2}\, ,
\end{equation}
where $I_{\mathrm{contr}}$ is the contracted integral evaluating to~\eqref{eq:icont1}.

To recover the local differential equation, it is necessary to first rewrite $Q_1 $ and $Q_3 $ in terms of $X_1$ and $X_2$. Following this, we will demonstrate that their product can be reformulated to depend solely on either $X_1$ or $X_2$, thereby recovering equation~\eqref{eq:locality}.

\paragraph{Reduction operators in terms of the physical variables.} 

To find the reduction operators in terms of $X_1$ and $X_2$, it is possible to use the relations between the $z_I$ and the physical variables specified in equation~\eqref{eq:physvars}. A simple application of the chain rule results in
\begin{equation}
        \frac{\partial}{\partial X_1}= \partial_1 +\partial_5  \, ,\quad
        \frac{\partial}{\partial X_2} = \partial_3 +\partial_5  \, ,\quad
         \frac{\partial}{\partial Y} = \partial_1 +\partial_3\, .
\end{equation}
However, these will not all be independent, because the Euler equations~\eqref{eq:exchangeeuler} give additional relations between the partial derivatives $\partial_I$. This makes it possible to eliminate derivatives of $Y$ by replacing
\begin{equation}
    Y \frac{\partial}{\partial Y}\simeq_{\E+\nu} 2\epsilon -1 -X_1 \frac{\partial}{\partial X_1}-X_2\frac{\partial}{\partial X_2}\, ,
\end{equation}
where, as before, $\simeq_{\E+\nu}$ denotes that equality holds when considering the equivalence relations stemming from both the toric and the Euler equations.

With these replacements, it is possible to obtain the action of the reduction operators in terms of $X_1$ and $X_2$. $Q_1 $ and $Q_3 $ then take the form
\begin{equation}\label{eq:Q13phys}
\begin{array}{rl}
     Q_1 \big\vert_{\mathrm{phys}}&\simeq_{\E+\nu} (X_1+Y)\frac{\partial}{\partial X_1} -\epsilon \eqqcolon Q_{1,X} \, , \\
      Q_3 \big\vert_{\mathrm{phys}}&\simeq_{\E+\nu} (X_2+Y)\frac{\partial}{\partial X_2} -\epsilon \eqqcolon Q_{3,X}\, .
\end{array}
\end{equation}
Therefore, equation~\eqref{eq:Q13nu=ict} can be rewritten as
\begin{equation}\label{eq:q13XI}
\left((X_2+Y)\frac{\partial}{\partial X_2} -\epsilon\right)\left((X_1+Y)\frac{\partial}{\partial X_1} -\epsilon\right)I=\frac{\partial^2 I_{\mathrm{contr}}}{\partial X_1^2}    \, ,
\end{equation}
where $I$ is the original single exchange integral~\eqref{eq:singleexchange}. Note that since the single exchange integral only depends on $X_1$ and $X_2$ through the combination $X_1+X_2$ it is already possible to rewrite this into a differential equation only involving one parameter, and this approach will lead to the locality differential equation~\eqref{eq:locality}. However, it turns out that this property is also encoded in the reduction operators themselves, due to a particular property satisfied by all reduction operators.

\paragraph{Reduction of variables and locality.} The key insight is that all reduction operators have the property that
\begin{equation}
    \partial_I Q_I (\nu)f(z;\nu)=0\ ,
\end{equation}
for all solutions $f$ at a parameter $\nu$. This makes it possible to eliminate derivatives of $\partial_I$ when acting on $Q_I (\nu)f$. In particular, it is possible to rewrite 
\begin{equation}
    \partial_1 Q_{1,X} \simeq_{\E+\nu} \frac{1}{2Y}\left(2\epsilon-1 +(Y-X_1)\frac{\partial}{\partial X_1}-(Y+X_2)\frac{\partial}{\partial X_2}\right)Q_{1,X}\, ,
\end{equation}
which can be used to eliminate the $X_2$ derivative in equation~\eqref{eq:q13XI}. After this replacement, equation~\eqref{eq:q13XI} can be rewritten as
\begin{equation}
    \left((X_1-Y)\frac{\partial}{\partial X_1}+\epsilon-1\right)\left((X_1+Y)\frac{\partial}{\partial X_1} -\epsilon\right) I=\frac{\partial^2 I_{\mathrm{contr}}}{\partial X_1^2} \, ,
\end{equation}
which, when expanded, recovers equation~\eqref{eq:locality}. Note that, because $Q_{1,X}$ and $Q_{3,X}$ commute, this procedure also makes it possible to obtain a differential equation involving only $X_2$.

\paragraph{Reduction operators and boundary conditions.}

Interestingly, the reduction operators and the resulting inhomogeneous differential equations can also be used to obtain boundary conditions for the single-exchange integral. In particular, consider $Q_1I_{\R^2_+}(z;\nu)\vert_{\rm phys}$ as in equation~\eqref{eq:Q1action} and let us write $I_{\rm phys}=I_{\R^2_+}(z;\nu)\vert_{\rm phys}$ to lighten the notation somewhat. Then, we want to obtain boundary conditions for $I_{\rm phys}$, such that we can write it in terms of the most general solution of the GKZ system, given by
\begin{equation}\label{eq:gensolphys}
    I_{\rm phys}=c_1(\R^2;\epsilon)f_1\vert_{\rm phys}+c_2(\R^2;\epsilon)f_2\vert_{\rm phys}+c_3(\R^2;\epsilon)f_3\vert_{\rm phys}+c_4(\R^2;\epsilon)f_4\vert_{\rm phys}\,.
\end{equation}
 Here the functions $f_i\vert_{\rm phys}$ are as written in equation~\eqref{eq:gensols} and the coefficients $c_i$ are what we are trying to solve for. 
 
 The first approach to obtaining these coefficients is that one can simply act on this general solution with a reduction operator. Using the inhomogeneous differential equation from~\eqref{eq:Q1action}, this will fix some of the coefficients $c_i$. In particular, acting with $Q_{1,X}$ on $I_{\rm phys}$ as above we find
 \begin{equation}
 \begin{split}
     Q_{1,X}I_{\rm phys} = &\epsilon \,c_3(\R_+^2;\epsilon)(X_1-Y)^{\epsilon-1}(X_2+Y)^\epsilon\\
     & +\epsilon\,c_4(\R_+^2;\epsilon)\frac{(X_1+X_2)^{2\epsilon}}{X_1-Y}\left(\, _2F_1\left(1,-2 \epsilon ;1-\epsilon ;\frac{X_2-Y}{X_1+X_2}\right)-1\right)\,,
\end{split}
 \end{equation}
where $_2F_1$ is again the hypergeometric function. Note that this equation no longer depends on $c_1$ and $c_2$ as $f_1\vert_{\rm phys}$ and $f_2\vert_{\rm phys}$ are annihilated by $Q_{1,X}$. Now, we must impose that this equation is equal to the integral on the right-hand side of equation~\eqref{eq:Q1action}. Evaluating that integral and solving for the coefficients $c_i$ we find
\begin{equation}
    c_3(\R_+^2;\epsilon)=0\, , \quad c_4(\R_+^2;\epsilon)= 2^{-2 \epsilon -1} \sqrt{\pi } \csc (\pi  \epsilon ) \Gamma \left(\frac{1}{2}-\epsilon \right) \Gamma (\epsilon )
\end{equation}
in accordance with equation~\eqref{eq:coefssols}.

A second approach to solving for the coefficients in equation~\eqref{eq:gensolphys} comes from the observation that, for $\epsilon>0$, all the solutions are regular in the limit $X_1+Y\rightarrow 0$\,.\footnote{In fact, this approach also works for general $\epsilon$ as both sides must obey the same scaling behavior in $X_1+Y$.} Therefore, using the explicit expression of $Q_{1,X}$ from equation~\eqref{eq:Q13phys}, we find that 
\begin{equation}
    Q_{1,X}I_{\rm phys}\vert_{X_1+Y\rightarrow 0}=\left((X_1+Y)\frac{\partial I_{\rm phys}}{\partial X_1}-\epsilon I_{\rm phys}\right)\bigg\vert_{X_1+Y\rightarrow 0}=-\epsilon I_{\rm phys}\vert_{X_1+Y\rightarrow 0}
\end{equation}
Using this observation and using the inhomogeneous equation for $Q_{1,X}$, we find that
\begin{equation}
    -\epsilon I_{\rm phys}\vert_{X1\rightarrow -Y} =  -\epsilon \int_{\R_+^2}d\omega_1 d\omega_2\; \frac{(\omega_1 \omega_2)^\epsilon}{(\omega_2+X_2+Y)(\omega_1+\omega_2+X_2-Y)^2}\,.
\end{equation}
Evaluating the integral on the right-hand side and taking the limit of the functions $f_i$, we would obtain the same coefficients as before.

\paragraph{Locality and twists.} As we have seen, the existence of these local differential equations requires the existence of both $Q_1 $ and $Q_3 $. Since the existence of these operators depends on the reducibility of the GKZ system this provides us with a way of classifying when these local differential equations exist. In fact, the reducibility of the GKZ system is fully encoded in the twist parameter $\nu$. It follows from the discussion in section~\ref{sec:reductions} that, if there exists a resonant face $F$ with $1 \not\in F$, then the reduction operator $Q_1 $ exists for suitable $\nu$.\footnote{Here, suitable means that $\nu$ is not in the $\C$-span of $F$, while $\nu-a_1$ is in the $\C$-span of $F$. Note that, if this is not the case, it is possible to parameter shift the integral by applying certain differential operators, as described in section~\ref{ssec:reductionoperators}.} Careful analysis then shows that this is the case only if $\nu_1$ is integer. Where we recall that the parameter $\nu$ encodes the twists of the integral
\begin{equation}
     I(X;\nu)\coloneqq \int_{R_+^2}\frac{\omega_1^{\nu_4-1} \omega_2^{\nu_5-1}\;d\omega_1 d\omega_2}{(\omega_1+X_1+Y)^{\nu_1}(\omega_2+X_2+Y)^{\nu_2}(\omega_1+\omega_2+X_1+X_2)^{\nu_3}}\, .
\end{equation}
A similar analysis for $Q_3 $ then implies that $\nu_2$ must also be integer. 

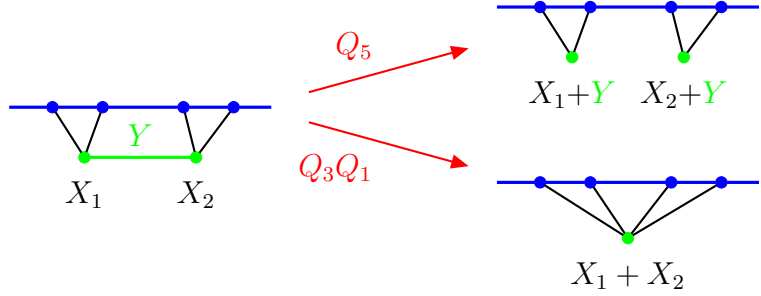
\begin{figure}
\centering
    \begin{tikzpicture}
        \begin{feynman}
            \vertex (topleft);
            \vertex [right=.5 cm of topleft,boundarydot] (topcenter1) {};
            \vertex [right=0.667 cm of topcenter1,boundarydot] (topcenter2) {};
            \vertex [right=.5 cm of topcenter2,boundarydot] (topcenter);
            \vertex [below=0.667 cm of topcenter] (midcenter) ;
            \vertex[left=0.667 cm of midcenter,bulkdot] (leftdiagram) {};
            \vertex[right=0.667 cm of midcenter,bulkdot] (rightdiagram){};
            \vertex [right=.5 cm of topcenter,boundarydot] (topcenter3) {};
            \vertex [right=0.667 cm of topcenter3,boundarydot] (topcenter4) {};
            \vertex [right =.5 cm of topcenter4] (topright);

            \vertex [right=3 cm of topright] (basepoint);

            \vertex [above=1.333 cm of basepoint] (topleftcut);
            \vertex [right=.5 cm of topleftcut,boundarydot] (topcenter1cut) {};
            \vertex [right=0.667 cm of topcenter1cut,boundarydot] (topcenter2cut) {};
            \vertex [right=.5 cm of topcenter2cut,boundarydot] (topcentercut);
            \vertex [below=0.667 cm of topcentercut] (midcentercut) ;
            \vertex[left=0.667 cm of midcentercut,bulkdot] (leftdiagramcut) {};
            \vertex[right=0.667 cm of midcentercut,bulkdot] (rightdiagramcut){};
            \vertex [right=.5 cm of topcentercut,boundarydot] (topcenter3cut) {};
            \vertex [right=0.667 cm of topcenter3cut,boundarydot] (topcenter4cut) {};
            \vertex [right =.5 cm of topcenter4cut] (toprightcut);

            \vertex [below=1cm of basepoint] (topleftcont);
            \vertex [right=.5 cm of topleftcont,boundarydot] (topcenter1cont) {};
            \vertex [right=0.667 cm of topcenter1cont,boundarydot] (topcenter2cont) {};
            \vertex [right=.5 cm of topcenter2cont,boundarydot] (topcentercont);
            \vertex [below=.667 cm of topcentercont,bulkdot] (midcentercont) {};
            \vertex [right=.5 cm of topcentercont,boundarydot] (topcenter3cont) {};
            \vertex [right=0.667 cm of topcenter3cont,boundarydot] (topcenter4cont) {};
            \vertex [right =.5 cm of topcenter4cont] (toprightcont);

            \vertex[right=.5cm of topright] (arrowanchor);
            \vertex[right=2cm of arrowanchor] (arrowanchor2);
            \vertex[above=.2cm of arrowanchor] (arrowbasetop);
            \vertex[above=.75cm of arrowanchor2] (arrowheadtop);
            \vertex[below=.2cm of arrowanchor] (arrowbasebot);
            \vertex[below=.75 cm of arrowanchor2] (arrowheadbot);            
            
            \diagram*  {
              (topleft) --[very thick,blue] (topcenter1) --[very thick,blue] (topcenter2) --[very thick,blue] (topcenter3) --[very thick,blue] (topcenter4) --[very thick,blue] (topright);
              (leftdiagram) --[thick,opacity=1] (topcenter1);
              (leftdiagram) --[thick,opacity=1]  (topcenter2);
              (rightdiagram) --[thick,opacity=1] (topcenter3);
              (rightdiagram) --[thick,opacity=1] (topcenter4);
              (leftdiagram) --[green,very thick] (rightdiagram);

              (topleftcut) --[very thick,blue] (topcenter1cut) --[very thick,blue] (topcenter2cut) --[very thick,blue] (topcenter3cut) --[very thick,blue] (topcenter4cut) --[very thick,blue] (toprightcut);
              (leftdiagramcut) --[thick,opacity=1] (topcenter1cut);
              (leftdiagramcut) --[thick,opacity=1]  (topcenter2cut);
              (rightdiagramcut) --[thick,opacity=1] (topcenter3cut);
              (rightdiagramcut) --[thick,opacity=1] (topcenter4cut);

              (topleftcont) --[very thick,blue] (topcenter1cont) --[very thick,blue] (topcenter2cont) --[very thick,blue] (topcenter3cont) --[very thick,blue] (topcenter4cont) --[very thick,blue] (toprightcont);
              (midcentercont) --[thick,opacity=1] (topcenter1cont);
              (midcentercont) --[thick,opacity=1]  (topcenter2cont);
              (midcentercont) --[thick,opacity=1] (topcenter3cont);
              (midcentercont) --[thick,opacity=1] (topcenter4cont);

              (arrowbasetop) --[edge label=$Q_5 $, thick,red,with arrow =1]  (arrowheadtop);
              (arrowbasebot) --[edge label'=$Q_3 Q_1$, thick,red,with arrow =1]  (arrowheadbot);
            };
            \vertex[below=.5 cm of leftdiagram] (X1) {\textbf{$X_1$}};
            \vertex[below=.5 cm of rightdiagram] (X2) {\textbf{$X_2$}};
            \vertex[above=0cm of midcenter] (Y) {\textbf{\color{green} $Y$}};

            \vertex[below=.5 cm of leftdiagramcut] (X1cut) {\textbf{$X_1 +$\textbf{\color{green} $Y$}}};
            \vertex[below=.5 cm of rightdiagramcut] (X2cut) {\textbf{$X_2 +$\textbf{\color{green} $Y$}}};

            \vertex[below=.5cm of midcentercont] (x1x2cont) {\textbf{$X_1+X_2$}};
        \end{feynman}
    \end{tikzpicture}
    \caption{Acting with the reduction operators makes it possible to either contract or remove an edge from the diagram.}\label{fig:Qedgereduction}
\end{figure}

This partially recovers a result from \cite{arkani-hamed_differential_2023} where the locality of the single-exchange integral was also studied. However, there it was found that $\nu_3$ must also be integer, implying that none of the polynomials in the numeral can be twisted. From the perspective of the reduction operators, this implies that $Q_5 $ exists and that the single exchange integral satisfies an equation of the type~\eqref{eq:Q5action}. Interestingly, the diagram corresponding to this equation is the ``cut" diagram of figure~\ref{eq:singleexchangecut}. Therefore, it is possible to obtain differential operators for the single exchange integral that, diagrammatically, correspond to either edge contraction or edge cutting, as shown in figure~\ref{fig:Qedgereduction}. Furthermore, locality is equivalent to both of these reductions being possible.

One is naturally led to wonder if this story can be generalized, especially since the reduction operators can be obtained for any reducible GKZ system. Since these systems include all cosmological correlators, it might be possible to perform these diagrammatic reductions by applying the reduction operators. Furthermore, one can try to assign a complexity to a diagram or integral by studying how many of these differential operators (and of what order) are necessary. We leave this and other such inquiries to further work \cite{futureworkGHV}.

\section{Conclusions}

This work was motivated by the goal of efficiently evaluating physically relevant integrals that solve systems of differential equations. We focused on systems that permit reductions to smaller subsystems, which can be determined algorithmically. To formalize this, we centered our attention on GKZ systems, which provide a well-defined and broad framework for studying these reductions in full generality. GKZ systems are constructed from a simple set of data —a matrix $\A$ and a vector $\nu$— which together define the differential equations and candidate solutions expressed as integrals. We provided the necessary and sufficient conditions that $\A$ and $\nu$ must satisfy for a GKZ system to reduce to smaller subsystems. Additionally, we introduced a general algorithm that determines the reduction operators required to explicitly perform the reduction and obtain the solution spaces of the subsystems.

To illustrate the reduction algorithm, we analyzed the GKZ system associated with a single exchange Feynman integral relevant to a specific cosmological correlator. In this example, we successfully determined all the reduction operators and obtained a complete set of solutions for the smaller subsystems. Even this basic example demonstrated the potential of this approach to lead to significant simplifications, as the step-by-step procedure circumvented the need to find the full four-dimensional solution space at once. Moreover, the subsystems involved differential equations that were far less complex than those of the full GKZ system, allowing us to obtain their associated solutions in closed form. 

One strength of the presented approach lies in its generality and algorithmic nature. A vast class of integrals of physical interest can be described as solutions to GKZ systems and in the majority of cases these systems are expected to be reducible. 
Deriving the reduction operators and their solution spaces for additional examples, such as other cosmological correlators or more general Feynman integrals, would be a valuable next step.  
Additionally, our detailed example suggested that the existence of reduction operators is tied to a physical property of the underlying theory -- the locality of the theory. 
This property was recently linked to the twists of the single exchange integral \cite{arkani-hamed_differential_2023}, and the existence of additional differential operators acting naturally on the physical solutions. We demonstrated in section~\ref{ssec:physics} that these operators are precisely the reduction operators and showed in section~\ref{sec:reductions} that the relationship between twists and reduction operators persists for general GKZ systems. This indicates that a remarkable connection between the reducability of differential equations and the locality of the theory persists for all cosmological correlators. We hope to elaborate on this in the future \cite{futureworkGHV}.

We expect that the reduction techniques are generally useful in the study of Feynman integrals. Considering them in their Feynman or Lee-Pomeransky representation \cite{weinzierl_Feynman_2022}, one shows that these integrals can be interpreted as GKZ systems and that most of the parameter vector $\nu$ will consist of integers. Furthermore, there are known reduction formulas for the polynomials in these integrals from a graph theoretic perspective \cite{bogner_feynman_2010}. Therefore, it seems likely that possible reductions for the polynomials in the integral will lead to reductions for the underlying GKZ system. If this is possible it could greatly simplify solving certain Feynman integrals and hopefully allow for systematic studies of classes of Feynman integrals, similar to those performed in \cite{bonisch_analytic_2021}.

Another interesting future application arises when considering integrals in a more geometric setting. It has been shown that period integrals of algebraic varieties are solutions to special GKZ systems \cite{hosono_mirror_1995,hosono_gkzCY_1996,hosono_gkzapp_1996}.These GKZ systems often have a larger solution space than expected from geometry. Interestingly though, it turns out that the resulting set of differential equations factorizes, and the identification of suitable factors results in the correct set of differential equations -- the Picard-Fuchs equations \cite{hosono_mirror_1995,hosono_gkzCY_1996}. This factorization is a consequence of the reducibility of the underlying GKZ system and the tools introduced in this paper can be readily applied to finding the geometric differential operators. For example, for the quintic hypersurface in $\P^4$ one easily obtains the actual Picard-Fuchs equations from the GKZ system by finding an appropriate reduction operator. One can expect that further reductions arise if the underlying variety has special additional features, such as a fibration structure, and it would be interesting to explore this further. The geometric examples are particularly interesting since, for period integrals, the parameter $\nu$ will consist only of integers and therefore will be as resonant as possible.

A main lesson we draw from our investigation is that the reduction process always leads to a reduction in complexity of finding a solution. This is true in a colloquial sense, since after the reduction the solutions obey a lower-order differential equations and a complicated GKZ system gets mapped to multiple smaller and simpler systems. Beyond this intuitive simplifications, we can also aim to \textit{quantify} 
this reduction of complexity in a precise way. In order to do that a general notion of complexity of physical quantities needs to be established. It was suggested in \cite{grimm_complexity_2024} that this can indeed be done using the major mathematical advances of \cite{binyamini_sharply_2022,binyamini_tameness_2023} on 
complexity and tameness. 
These techniques allow for assigning a measure of complexity to functions, or more general sets, and thus give a natural measure of the complexity or information content of a physical observable. It was shown in \cite{grimm_complexity_2024} that the complexity of all tree-level cosmological correlators can be computed explicitly. Furthermore, it was
put forward in \cite{grimm_complexity_2024,grimm_structure_2024} that in general all Feynman integrals should have a notion of complexity, since they are period integrals and tame functions \cite{douglas_tameness_2022}. With these advanced one can make the decrease in complexity due to the existence of a reducible GKZ system precise as we will see in an upcoming work \cite{futureworkGHV}. Combining this with the relation between reducibility and locality, we are led to the remarkable conclusion that the locality of the theory reduces the computational complexity in quantifiable way. It was speculated in \cite{grimm_structure_2024} that the computational complexity should be identified with cosmological time. This implies that any reduction will have far-reaching implications and, as noted already in \cite{grimm_structure_2024}, might help to solidify the identification of cosmological time and computational complexity. 

\subsubsection*{Acknowledgements}

We profited immensely from discussions with Damian van de Heisteeg, Andreas Hohl, Jeroen Monnee, Guilherme Pimentel, Anna-Laura Sattelberger, and Mick van Vliet. 
This research is supported, in part, by the Dutch Research Council (NWO) via a Vici grant.

\appendix

\section{GKZ systems and their power series solutions} \label{ap:GKZ}

In this appendix we will review some of the general properties of GKZ systems. These are well-known and many of these are either from the original papers by Gelfand, Kapranov, and Zelevinsky \cite{gelfand_generalized_1990,gelfand_hypergeometric_1991,gelfand_discriminants_1994} or are from the excellent book \cite{saito_Grobner_2000}. The presentation and results described here will follow this last reference, as well as some more modern references that focus specifically on Feynman integrals \cite{nasrollahpoursamami_periods_2016,delacruz_feynman_2019,blumlein_hypergeometric_2021,klausen_hypergeometric_2020}. We will continue from the description of general GKZ systems in section~\ref{ssec:gengkz}.

\subsection{Differential equations and hypergeometric integrals.} 
\label{ap:difeqs}

We begin by describing the relation between the system of differential equations and the hypergeometric integrals. Recall from \eqref{eq:gengkzint} that these integrals were of the form
\begin{equation}\label{eq:apintegral}
   I(z;\alpha,\beta)= \int_\Gamma d^n\omega\; \frac{\prod_{i=1}^n \omega_i^{\alpha_i-1}}{\prod_{j=1}^k p_j(z,\omega)^{\beta_j}}\, ,
\end{equation}
where $\Gamma$ is an arbitrary integration cycle, $\alpha_i$ and $\beta_j$ are complex numbers and the $p_j$ are polynomials in the $\omega_i$ with coefficients $z_I$. 

For convenience, we will introduce a bit of notation. For each $l$-dimensional vector and list of $l$ objects $c_i$, we will write
\begin{equation}
    c^v \coloneqq \prod_{i=1}^l c_i^{v_i}\, .
\end{equation}
Note that we have already introduced this notation for the partial derivatives $\partial^u$ in the main text. With this notation, the integral~\eqref{eq:apintegral} can be written as
\begin{equation}
     I(z;\alpha,\beta)= \int_\Gamma d^n\omega\; \frac{\omega^{\alpha-1}}{p(z,\omega)^\beta}\, .
\end{equation}
Recall that the matrix $\A$ was constructed from a set of partial matrices $\A_j$ for each polynomial $p_j$, with a homogenization step that fills the top $k$ rows with ones and zeroes. These components of 
$\A$ will exhibit different behaviors when considering the differential equations of this GKZ system. Therefore, we will refer to the top $k$ rows as the homogeneous part and the bottom $n$ rows as the polynomial part. We will denote these by $\A_\beta$ and $\A_\alpha$, respectively. Schematically, this split takes the form
\begin{equation}
    \A\coloneqq \begin{pmatrix}
    \mathbf{1} & \mathbf{0} &\cdots\\
    \mathbf{0} & \mathbf{1} & \cdots \\
    \vdots&\ddots&\ddots\\
    \A_1 & \A_2 &\cdots 
    \end{pmatrix}
    \begin{array}{l}
         \left\}\begin{array}{l}
              \, \\
            \A_\beta    \\
              \,
         \end{array} \right.\\
          \left\}\begin{array}{l}
           \; \A_\alpha
         \end{array} \right.
    \end{array}
\end{equation}
where $\mathbf{1}$ and $\mathbf{0}$ denote row vectors consisting of either ones or zeroes. 

Similarly, we can separate each vector $a_I$ into its first $k$ components $a_{I,\alpha}$ and its last $n$ components $a_{I,\beta}$, such that $a_I=(a_{I,\beta},a_{I,\alpha})$, similar to how we split $\nu=(\beta,\alpha)$. Note that $a_{I,\alpha}$ is the exponent of the term
\begin{equation}\label{eq:aialpha}
  p_j(z,\omega)=  \cdots+z_I \omega^{a_{I,\alpha}}+\cdots\ 
\end{equation}
of the polynomial $p_j$, while $a_{I,\beta}$ is the unit vector in the $j$-th direction. Therefore, one can interpret $a_{I,\beta}$ as indicating the polynomial in which $z_I$ appears, while $a_{I,\alpha}$ specifies the exponent of $\omega$ with which it is associated.

\paragraph{The toric equations.} 

We will begin by showing that the toric operators annihilate the integral, regardless of the integration cycle. Acting with a partial derivative $\partial_I$ on $p^{-\beta}$ results in
\begin{equation}\label{eq:dzjm}
    \partial_I p(z;\omega)^{-\beta} = -(\beta\cdot a_{I,\beta})\, \omega^{a_{I,\alpha}}\, p(z,\omega)^{-\beta-a_{I,\beta}}\, ,
\end{equation}
where we recall that $a_I=(a_{I,\beta},a_{I,\alpha})$ and $\cdot$ denotes the vector dot product.

Similarly, for a vector $u$ with positive integer entries, one can act with $\partial^u$ on $p^{-\beta}$ yielding
\begin{equation}
    \partial^u p(z,\omega)^{-\beta}=c_u(\beta)\,\omega^{\A_\alpha u} \,p(z,\omega)^{-\beta-\A_\beta u}\, ,
\end{equation}
where $c_u(\beta)$ arises from an iterative application of equation~\eqref{eq:dzjm}. Note that $\A u=\A v$ if and only if $\A_\alpha u=\A_\alpha v$ and $\A_\beta u=\A_\beta v$. Furthermore, careful evaluating of the pre-factor $c_u(\beta)$ reveals that it depends only on $\A_\beta u$. Therefore, if $\A u=\A v$ for two vectors with positive integer entries, it follows that
\begin{equation}
    (\partial^u-\partial^v) p(z,\omega)^{-\beta} =0\, .
\end{equation}
Since these derivatives commute with the integral with respect to $\omega$, this gives rise to the toric equations.

\paragraph{The Euler equations.} Recall that the Euler operators arise from multiplying the matrix $\A$ with a vector of homogenous derivatives 
\begin{equation}
    \Theta=(\theta_I)_{1\leq I \leq N}=(z_J \partial_I)_{1\leq I\leq N}\, .
\end{equation}
Here as well, it is beneficial to separate the matrix $\A$ into its homogeneous part and the rest of $\A$. Specifically, we will first examine the Euler equations
\begin{equation}
    (\E_J+\nu_J)I(z;\alpha,\beta)=0
\end{equation}
with $J>k$, where $k$ is the number of polynomials $p_j$. These equations correspond to the homogeneous part.  Subsequently, we will consider the Euler equations for $J \leq k$.  Note that $\nu_J=\beta_J$ for $J\leq k$ and $\nu_J=\alpha_{J-k}$ for $J>k$.

Let us begin by considering the integral~\eqref{eq:apintegral} and perform a coordinate transformation $\omega_i\rightarrow s\, \omega_i$, with $s\in \C$. We assume that, for $s$ sufficiently close to $1$, the contour $\Gamma$ is invariant under this transformation. Considering the image of $p(z,\omega)$ under this transformation, one finds
\begin{equation}
    p(z,\omega)\rightarrow p\left(z_1 s^{(a_{\alpha,1})_i},z_2  s^{(a_{\alpha,2})_i},\cdots,z_N s^{(a_{\alpha,N})_i},\omega \right)\, .
\end{equation}
In other words, the transformation can be negated by a suitable inverse transformation of the $z_I$. It follows that, in total, applying this coordinate transformation to the integral~\eqref{eq:apintegral} has the effect
\begin{equation}
     I(z;\alpha,\beta)= \int_\Gamma d^n\omega\; \frac{\omega^{\alpha-1}}{p(z,\omega)^\beta}=s^{\alpha_i} I\left(s^{(a_{\alpha,I})_i}z_I;\alpha,\beta\right)\, ,
\end{equation}
where the pre-factor is due to the transformation of $d^n\omega\, \omega^{\alpha-1}$. Differentiating both sides with respect to $s$ and taking the limit $s\rightarrow 1$ results in
\begin{equation}
   0= \alpha_i I(z;\alpha,\beta)+\left(\sum_{I=1}^N (a_{\alpha,I})_i z_I \partial_I\right)I(z;\alpha,\beta)\, .
\end{equation}
Recognizing that 
\begin{equation}
    \sum_{I=1}^N (a_{\alpha,I})_i\, z_I \partial_I=\E_{i+k}\, ,
\end{equation}
with $k$ the number of polynomials $p_j$, we recover the Euler equations
\begin{equation}\label{eq:eulerderiv1}
    (\E_{i+k}+\nu_{i+k})I(z;\alpha,\beta)=0
\end{equation}
for $1\leq i\leq n$.

The other Euler equations can be obtained by noting that these correspond to the homogenous part $\A_\beta$ of $\A$. Therefore, these will only involve derivatives with respect to the variables contained in a single polynomial. In fact, one finds that, for $1\leq j \leq k$,
\begin{equation}
    \E_j\, p_j^{-\beta_j}=\left(\sum_{I\in p_j} z_I \partial_I \right)\left(\sum_{I\in p_j} z_I \omega^{a_I} \right)^{-\beta}=-\beta_j \,p_j^{-\beta_j}\, ,
\end{equation}
where the sum is only over the terms contained in $p_j$. Since this Euler operator acts trivially on all other $p_l$ for $l\neq j$, we obtain
\begin{equation}
    (\E_j+\nu_j)I(z;\alpha,\beta)=0
\end{equation}
for $1\leq j \leq k$. This, combined with equation~\eqref{eq:eulerderiv1}, yields all the Euler equations.

\subsection{General series solutions}
\label{ap:series}

One remarkable aspect of GKZ systems is that, provided certain technical conditions are met, their solutions can be counted in an entirely geometrical manner. Moreover, it is often possible to explicitly obtain the series expansions of these solutions, with the formulas being completely determined by the underlying geometric structure. This section is entirely based on \cite{saito_Grobner_2000} and references therein.  For a detailed description of these results and their proofs, we refer the reader to these sources.

\paragraph{Convex polytopes and number of solutions.}

The main geometrical structure underlying GKZ systems is a convex polytope that we associate to the matrix $\A$. To construct it, recall that we labeled the columns of $\A$ as $a_I$. Since this is a collection of vectors, one can take their convex hull and denote it as
\begin{equation}
    \mathrm{Conv}(\A)\coloneqq \mathrm{Conv}(a_1,a_2,\cdots,a_N)\, .
\end{equation}
Note that the resulting polytope can also be obtained from the Newton polytopes of the polynomials $p_j$.

Interestingly, important information about the number of solutions to the GKZ system can be derived from the volume of this polytope. In particular, only assuming that $\A$ is homogenized, one finds that \cite[Thm 3.5.1]{saito_Grobner_2000}
\begin{equation}\label{eq:solsandvol}
    \# \text{ of solutions }\geq \text{Vol}(\A)\, ,
\end{equation}
where $\text{Vol}(\A)$ is volume of the polytope $\mathrm{Conv}(\A)$, normalized such that the standard simplex has volume one. If we assume that $\A$ is normal, an assumption satisfied by most Feynman integrals \cite{tellander_cohenmacaulay_2023} as well as cosmological correlators, this inequality turns into an equality. Similarly, if the parameter $\nu$ is generic, meaning it is outside of the so-called exceptional hyperplane arrangement,\footnote{See \cite[Sec 4.5]{saito_Grobner_2000} for precise definitions.} equation~\eqref{eq:solsandvol} also becomes an equality.

\paragraph{Series expressions from triangulations.}

Besides the number of solutions, the polytope $\mathrm{Conv}(\A)$ also gives rise to a way of obtaining these solutions, again assuming that $\nu$ is generic. These solutions are associated to a triangulation of this polytope, and specifically the simplices of this triangulation.

If we let $\mathcal{T}$ to be a triangulation of $\mathrm{Conv}(\A)$, and assume that it is both unimodular, in the sense that each simplex has a normalized volume of one, and regular, in the sense of \cite[Ch 8]{sturmfels_Grobner_1997} each simplex will correspond to a single solution. 

The explicit series expansion of these solutions are known as the canonical series, or the $\Gamma$-series of the GKZ system. To describe these functions, it is first necessary to introduce some notation. Recall that for any subset $F\subset \{1,\cdots,N\}$, we have defined the matrix $\A_F$ as the matrix with column vectors $a_I$ for $I\in F$. Furthermore, we have denoted $\bar{F}$ as the complement of $F$ in $\{1,\cdots, N\}$. We will also denote the set of coordinates indexed by $I\in F$ as $z_F$. Finally, we will need the multivariate Pochhammer symbol
\begin{equation}
(a)_n = \prod_k \frac{\Gamma(a_k+n_k)}{\Gamma(a_k)}
\end{equation}
with $\Gamma$ the $\Gamma$-function\footnote{Not to be confused with the $\Gamma$-\textit{series} we are defining here. This $\Gamma(z)$ is the familiar generalization of the factorial.} as well as the multivariate factorial $n!\coloneqq \prod_k n_k!$.

With these definitions at hand, the canonical series solution of a simplex $\sigma$ of the unimodular regular triangulation $\mathcal{T}$ is defined as \cite{klausen_hypergeometric_2020}
\begin{equation}\label{eq:canonicalseries}
    \phi_\sigma(z;\nu) \coloneqq z_\sigma^{-\A_\sigma^{-1}\nu} \sum_{n\in \N^r} \frac{(\A_\sigma^{-1} \nu)_{\A^{-1}_\sigma \A_{\bar{\sigma}}n}}{n!} \frac{z_{\bar{\sigma}}^n}{(-z_\sigma)^{\A_\sigma^{-1} \A_{\bar{\sigma}}n}}\, ,
\end{equation}
with $r=\mathrm{dim}(\ker(\A))$. Note that the simplices will correspond to square invertible matrices $\A_\sigma$, making their inverses well-defined. Furthermore, the number of summation parameters is exactly the number of independent parameter $s_i$.

To illustrate equation~\eqref{eq:canonicalseries} with an example. Note that the single-exchange matrix from equation~\eqref{eq:exchangeA} has the unimodular regular triangulation of
\begin{equation}\label{eq:triang}
T=\{\{1, 3, 4, 5, 6\}, \{1, 2, 3, 4, 6\}, \{1, 4, 5, 6, 7\}, \{1, 2, 4, 6, 7\}\}\, .
\end{equation}
The associated canonical series for $\sigma_1\coloneqq \{1, 3, 4, 5, 6\} $ is given by
\begin{equation}
\begin{split}
\phi_1=\frac{z_3^{\nu_5}z_5^{\nu_4}}{z_1^{\nu_1} z_3^{\nu_2} z_4^{\nu_5}z_5^{\nu_3}z_6^{\nu_4}} \sum_{n_1,n_2=0}^\infty &\frac{(\nu_1)_{n_1}(\nu_2-\nu_5)_{-n_2}(\nu_5)_{n_2}(\nu_3-\nu_4)_{n_2-n_1}(\nu_4)_{n_1}}{n_1!n_2!}\\
&\left(-\frac{z_2 z_5}{z_1 z_6}\right)^{n_1} \left(-\frac{z_3 z_7}{z_4 z_5}\right)^{n_2}
\end{split}
\end{equation}
for generic $\nu$.\footnote{In fact, for the single exchange GKZ system, this equation holds for all $\nu\neq 0$.} Note that in this formula the expressions of the form $(\nu_1)_{n_1}$ refer to the ordinary Pochhammer symbol defined as
\begin{equation}
    (a)_n\coloneqq \frac{\Gamma(a+n)}{\Gamma(n)}\, .
\end{equation}
We want to emphasize that, even if these series expressions can be obtained and result in a basis of solutions to the GKZ system, complications can arise as the number of summation parameters grows. Since then these types of series become increasingly difficult to evaluate. Furthermore, analytical continuations of such series expansions can also pose problems in such situations.

\section{Reductions using Euler-Koszul homologies} \label{ap:derivation}

In this appendix we discuss the central mathematical results used in the main text. To do this we introduce the language of $\D$-modules and the use of the so-called Euler-Koszul homology. 
In order to keep our exposition concise, we will not present 
all details about these mathematical concepts and instead refer to interested reader to the existing literature on these subjects. In particular, references \cite{andres_Constructive_2010,brodmann_Notes_2018,sattelberger_dmodules_2019} give a general discussion of $\D$-modules, while the works  \cite{saito_Grobner_2000,cattani_three_2006,stienstra_gkz_2005,klausen_hypergeometric_2020,feng_gkzhypergeometric_2020,henn_dmodule_2024} provide an overview of how GKZ systems are related with $\D$-modules. Introductions to Euler-Koszul homologies can be found in  \cite{matusevich_homological_2004,reichelt_algebraic_2021}.

The appendix is structured as follows. We start in section~\ref{sap:dmods} with a brief introduction to $\D$-modules and describe the $\D$-module associated to a GKZ system. Afterwards we will introduce the framework of Euler-Koszul homologies in section~\ref{sap:eulerkoszul} and discuss their relation to GKZ systems. In section~\ref{sap:reducibility} we discuss how reducibility of a $\D$-module gives rise to `simpler' solutions associated to a submodule and relate this observation to known results about the reducibility of GKZ systems. In  section~\ref{sap:reductionoperators} we build upon these results to obtain additional submodules of GKZ system, where the associated solutions are annihilated by special operators -- the reduction operators.

\subsection{GKZ systems as $\D$-modules}\label{sap:dmods}

We start by defining the Weyl algebra $\D_N$ in $N$ coordinates $z_I$. This is obtained by considering the free algebra $A_{2N}$ in the $2N$ variables $z_I$ and $\partial_I$, modulo the commutation relations
\begin{equation}
    [z_I,z_J]=0,\quad [\partial_I,\partial_J]=0,\quad [\partial_I,z_J]=\delta_{I,J}\ ,
\end{equation}
where $\delta_{I,J}$ is the Kronecker delta. In other words, the Weyl algebra in $N$ variables is defined as the quotient
\begin{equation}
  \D_N \coloneqq A_{2N}/\big( [z_I,z_J]\sim 0,[\partial_I,\partial_J]\sim 0, [\partial_I,z_J]-\delta_{I,J}\sim 0 \big)\, .
\end{equation}
Note that if it is clear from the context we will drop the subscript $N$ and write $\D=\D_N$.

We can use the Weyl algebra to study differential equations in the following way. Consider a set of differential operators $P_i$, it is possible to define the left $\D$-ideal $\I$ generated by these operators as
\begin{equation}
    \I=\langle P_i\rangle_\D \coloneqq \sum_i \D \cdot P_i\ ,
\end{equation}
which allows us to define the $\D$-module
\begin{equation}
    \M=\D/\I\, .
\end{equation}
The $\D$-module homomorphisms of this module are then related to the solutions of
\begin{equation}\label{eq:pif=0}
    P_i f=0
\end{equation}
as follows. Consider the space of holomorphic functions $\O$ in $N$ variables on an open subset of $\C^N$, where this open subset lies outside of the singular locus of the differential equations. We then consider the $\D$-homomorphisms $\Hom_\D(\M,\O)$ and claim that these are in one to one correspondence with solutions of equation~\eqref{eq:pif=0}. This correspondence is as follows, for any such homomorphism $\phi \in \Hom_\D(\M,\O)$ we have that
\begin{equation}
    0=\phi(0)=\phi(P_i)=P_i\phi(1)\, .
\end{equation}
since $P_i=0$ as an element of $\M$. Therefore, $\phi(1)$ is an element of $\O$ that satisfies the equations~\eqref{eq:pif=0}. Conversely, for any $f$ satisfying equation~\eqref{eq:pif=0} we obtain a homomorphism by defining $\phi(1)=f$. Since such a homomorphism is completely determined by its action on the identity we obtain the required result. Because of this relation, we will move between the different perspectives if one is more useful than the other. We will call such a $\phi$ or such an $f$ a \textit{solution of $\M$}, and write the vector space of such solutions as $\mathrm{Sol}(\M)$ where we have suppressed the dependence on $\O$.

In general, one often studies the sheafified versions of $\O$ and $\D$ over some algebraic variety. One then considers the solution complex $R \mathrm{Hom}_{\D_X}(\M,\O_X)$ in the derived category \cite{hotta_dmodules_2008}. Since we are only interested in obtaining the solutions around some point, we are free to take $X$ such that $\M$ is non-singular. In this case the homology of $R \mathrm{Hom}_{\D_X}(\M,\O_X)$ becomes concentrated in the zero-th degree and we obtain the solutions as described above. In particular, this implies that, when acting on such non-singular $\M$, the solution functor is exact.\footnote{We are grateful to Andreas Hohl and Anna-Laura Sattelberger for insightful discussions regarding this matter.}

It is natural to ask how many independent solutions there are for a given $\D$-module $\M$, or equivalently, what the dimension is of $\mathrm{Sol}(\M)$. This is known as the \textit{rank} of $\M$. For us, we will consider only so-called holonomic $\D$-modules, which implies that the rank is finite.

\paragraph{The GKZ module.}

We are now ready to rephrase the set of differential equations considered in section~\ref{sec:gkzsystems} in terms of $\D$-modules. We consider an $M$ by $N$ matrix $\A$ and an $M$-dimensional complex vector $\nu$. The first step is to use the differential equations to define an ideal. Recall that these come in two parts, we first consider the toric equations from equation~\eqref{eq:toricopdef}, leading to the ideal
\begin{equation}
    \I_\A \coloneqq \langle \L_{u,v} \; \vert \; \A u=\A v, \rangle_\D
\end{equation}
where $u,v$ are elements of $\N^M$. Similarly we can define an ideal by considering the Euler operators
\begin{equation}
    \langle \E_J+\nu_J\; \vert \;1\leq J \leq M\rangle_\D
\end{equation}
defined in equation~\eqref{eq:eulergen}. Combining these two ideals we obtain the GKZ ideal
\begin{equation}
    \H_\A(\nu)\coloneqq \I_\A+\langle \E_J+\nu_J\rangle_\D\, ,
\end{equation}
and with it, the GKZ module
\begin{equation}\label{eq:gkzmod}
\M_\A(\nu)\coloneq \D/\H_\A(\nu)\, ,
\end{equation}
which will be the $\D$-module whose solutions we want to obtain.

Similarly, for any subset $F\subset A\coloneqq \{1,\cdots,N\}$, we will write $z_F$, $\partial_F$ for the coordinates indexed by $F$. With this it is then possible to proceed along the same lines as before and define the Weyl algebra $\D_F$ for these coordinates. If we then define the matrix $\A_F$ by combining the column vectors of $\A$ for $i\in F$, we can consider the GKZ system this matrix defines. This GKZ system will then be a $\D_F$-module and we will write it as $\M_{\A_F}(\nu)$ for a parameter $\nu$. Note that by \cite[lemma 4.9]{matusevich_homological_2004}, $\cM_F(\nu)$ will have rank zero if $\nu$ is not in the $\C$-span of $F$.

\subsection{Euler-Koszul homologies}\label{sap:eulerkoszul}

While the $\D$-module $\M_\A(\nu)$ describes the GKZ system, it will be useful for us to consider a slightly different perspective as well. Instead of defining the GKZ system as above, we will consider it as the zero-th homology of a so-called Euler-Koszul complex \cite{matusevich_homological_2004,walther_duality_2005,matusevich_combinatorics_2004,berkesch_agraded_2009,schulze_hypergeometric_2009,berkesch_rank_2011,schulze_resonance_2012,reichelt_bFunctions_2018,steiner_ahypergeometric_2019,steiner_dualizing_2019,reichelt_algebraic_2021}. In this section we will briefly review its construction, referring to \cite{matusevich_homological_2004} for more details and proofs.

In order to define the Euler-Koszul complex, we first consider the commutative subring $R$ of $\D$, consisting only of the partial differentials:
\begin{equation}
   R\coloneqq \C[\partial_I]\simeq\D/\langle z_I\rangle_\D \ .
\end{equation}
 Since the toric operators involve only these partial differentials, it is also possible to consider the $R$-ideal generated by the toric operator $I_\A$ and use it to define the ring
\begin{equation}\label{eq:SAdef}
    S_\A\coloneqq R/I_\A\, .
\end{equation}
We will often switch between considering this quotient as a $\D$-module, as an $R$-module or as a ring itself, depending on the application. Furthermore, ideals of this ring will be written as $\langle \cdots \rangle$ without a subscript.

Note that $S_\A$ is naturally $\Z^N$ graded by defining the degree operator $\mathrm{deg}$ as
\begin{equation}
    \mathrm{deg}(\partial_I)=-a_I,\quad \mathrm{deg}(P_1 P_2)=\mathrm{deg}(P_1)+\mathrm{deg}(P_2)\, ,
\end{equation}
where $a_I$ is the $I$-th column of $\A$.\footnote{Note that the degree of an operator is a vector since the $a_I$ are vectors.} It is also possible to extend this grading to the full Weyl algebra $\cD$ by defining $\mathrm{deg}(z_I)=a_I$. We will call any module compatible with this grading to be \textit{$\A$ graded}. 

\paragraph{The Euler-Koszul complex.}

For any homogeneous operator $P_\alpha$ of degree $\alpha$, we can define the maps
\begin{equation}\label{eq:sjdef}
  s_J: P_\alpha \mapsto\ (\E_J+\nu_J-\alpha_J)P_\alpha\, ,
\end{equation}
where $\E_J$ is the $J$-th Euler operator defined in equation~\eqref{eq:eulergen} and $1\leq J\leq M$. These maps can then be linearly extended to non-homogeneous operators.  Furthermore, for a vector of operators $P \in (S_\A)^M$, we define
\begin{equation}
    s(P)= \sum_{J=1}^M s_J(P_J)
\end{equation}
where $P_J$ is the $J$-th element of $P$. With the map $s$, it is possible to define the Koszul Complex
\begin{equation}
   K_\bullet(\E+\nu):\quad 0\rightarrow \bigwedge^M \D^M\xrightarrow{d_M} \cdots\xrightarrow{d_1} \bigwedge^1 \D^M \xrightarrow{d_0} \D\rightarrow 0\, ,
\end{equation}
where the differentials are given by
\begin{equation}\label{eq:differentialdef}
    d_k(P_1\wedge\cdots\wedge P_k)=\sum_{i=1}^k (-1)^{i+1}s(P_i) \;P_1\wedge \cdots \wedge \widehat{P}_i\wedge \cdots \wedge P_k
\end{equation}
and $\widehat{P}_i$ means that this term is omitted. For any $\A$ graded $R$-module $S$ we consider it as a $\D$-module by taking $\mathcal{S}\coloneqq\D \otimes_{R} S$ and define the \textit{Euler-Koszul complex}
\begin{equation}
    K_\bullet(\E+\nu,S)\coloneqq K_\bullet(\E+\nu) \otimes_{\D} \mathcal{S}\, ,
\end{equation}
where the differentials are induced by the differential~\eqref{eq:differentialdef}. The \textit{Euler-Koszul homology} $H_i(\E+\nu,S)$ is then the $i$-th homology of this complex. Note that the zero-th homology $H_0(\E+\nu,S_\A)$ recovers the GKZ system from equation~\eqref{eq:gkzmod}.

\paragraph{Induced exact sequences.}

One useful property of the Euler-Koszul homologies is that a short exact sequence of $\A$ graded $R$-modules
\begin{equation}
    0\rightarrow S_1\rightarrow S_2\rightarrow S_3 \rightarrow 0
\end{equation}
with homogeneous maps will induce a long exact sequence of the form
\begin{equation}
\begin{array}{rll}
    \cdots &\rightarrow H_{i+1}(\E+\nu,S_3)&\rightarrow H_i(\E+\nu,S_1)\rightarrow H_i(\E+\nu,S_2)\\
    &\rightarrow H_i(\E+\nu,S_3) &\rightarrow\quad\cdots
\end{array}
\end{equation}
on the Euler-Koszul homologies. One particularly useful example of this is the exact sequence
\begin{equation}\label{eq:dIseq}
    0\rightarrow S_\A(a_I)\xrightarrow{\cdot \partial_I} S_\A \rightarrow S_\A/\langle \partial_I\rangle \rightarrow 0\ ,
\end{equation}
where $S_\A(a_I)$ is the module $S_\A$ with the degrees shifted by $a_I$ and $\cdot \partial_I$ denotes right multiplication with $\partial _I$. This particular sequence was used extensively in \cite{walther_duality_2005} and will play a major role in the results obtained in this paper. 

The exact sequence~\eqref{eq:dIseq} becomes especially useful when one considers that, for an $\A$ graded $R$-module $S$ and a vector $\alpha\in \C^M$, we can define its twist $S(\alpha)$ obtained by shifting the degrees of each operator with $\alpha$. Since the maps $s_J$ from equation~\eqref{eq:sjdef} are sensitive to this shift, we find
\begin{equation}
H_i(\E+\nu,S(\alpha))=H_i(\E+\nu+\alpha,S)\, .
\end{equation}
Therefore the sequence in equation~\eqref{eq:dIseq} results in the long exact sequence
\begin{equation}
\begin{array}{rll}
    \cdots&\rightarrow H_1(\E+\nu,S_\A/\langle \partial_I\rangle)&\rightarrow H_0(\E+\nu+a_I,S_\A)\xrightarrow{\cdot \partial_I} H_0(\E+\nu,S_\A)\\
    & \rightarrow H_0(\E+\nu,S_\A/\langle \partial_I\rangle)&\rightarrow 0
\end{array}
\end{equation}
which will be one of the key sequences used in this paper. Note that if $S_\A$ is a Cohen-Macaulay ring, all terms to the left of $H_1(\E+\nu,S_\A/\langle \partial_I\rangle)$ will be zero \cite[Remark 6.4]{matusevich_homological_2004}. This happens in many examples of interest, see for example the recent discussion in \cite{tellander_cohenmacaulay_2023}.\footnote{For a Cohen-Macaulay subring of $S_\A$, it is not guaranteed that the higher homology groups are zero since it may not be of maximal dimension.}

\subsection{Reducibility of GKZ systems and solution spaces}\label{sap:reducibility}
\label{assec:reducibility}
In this section we want to briefly explain how the existence of submodules leads to the ability to obtain a partial basis of the solution space. Afterwards we will apply this observation to a specific subsystem found in \cite{schulze_resonance_2012} and show how this subsystem manifests itself in terms of the solutions to the GKZ system.

\paragraph{Submodules and solution spaces.}

The crucial observation is that a submodule gives rise to a subset of the solutions satisfying additional of differential equations. This can be seen most clearly when considering solutions as $\D$-module homomorphisms. In this case,  any surjective $\D$-module homomorphism
\begin{equation}
    \Phi:\M\twoheadrightarrow\mathcal{N}
\end{equation}
induces an injective map on the solution spaces
\begin{equation}\label{eq:inducedsolmap}
    \Phi^*:\mathrm{Sol}(\mathcal{N})\hookrightarrow \mathrm{Sol}(\M)
\end{equation}
given simply by precomposition with $\Phi$. 

The solutions of $\M$ in the image of this map have some interesting properties. For any differential operator $P\in \ker(\Phi)$ and $\phi\in \mathrm{Sol}(\mathcal{N})$, we have that
\begin{equation}
   P\;\Phi^*(\phi)(1)= \Phi^*(\phi)(P)=0\, .
\end{equation}
In other words, the solutions in the image of $\Phi^*$ are exactly those solutions of $\M$ that are also annihilated by every $P\in \ker(\Phi)$. Therefore, if this kernel is non-trivial, there will be a part of the solution space that satisfies additional differential equations with respect to a general solution of $\M$. We now want to apply this perspective to GKZ systems.

\paragraph{Resonance and submodules.}
 
 In \cite{schulze_resonance_2012} the reducibility of a GKZ system was characterized in terms of its resonance. Furthermore, in the proof of \cite[Thm. 4.1]{schulze_resonance_2012} an explicit submodule was found, on the condition that there exists a non-trivial face $F$ such that $\nu$ is in the $\C$-span of $F$ and $\A$ is not a pyramid over $F$. Defining
\begin{equation}
    \bar{F}\coloneqq A \setminus F\,
\end{equation}
the natural surjection
\begin{equation}
    S_\A\twoheadrightarrow S_{\A_F}\simeq S_\A/ \langle \partial_{\bar{F}}\rangle\, ,
\end{equation}
induces a surjection 
\begin{equation}\label{eq:Fsurjection}
    H_0(\E+\nu,S_\A)\twoheadrightarrow H_0(\E+\nu,S_{\A_F})
\end{equation}
on the Euler-Koszul homologies, where we recall that $A\coloneqq \{1,\cdots,N\}$ and $\langle \partial_{\bar{F}}\rangle$ is the ideal generated by the partial derivatives not in $F$. Thus we find a surjective $\D$-module morphism as required to obtain relate the solution spaces as in~\eqref{eq:inducedsolmap}. 

Now let us consider some properties of this map and the solutions in its image. Clearly, the kernel of~\eqref{eq:Fsurjection} is generated by $\partial_I$ with $I\in A\setminus F$. Therefore, the associated solutions are exactly those with
\begin{equation}\label{eq:pif=02}
    \partial_I f=0
\end{equation}
for $I\in A\setminus F$. This provides us with the first set of subsystems described in section~\ref{ssec:reductionoperators}. 

There is an alternative characterization of the solutions associated to this submodule as solutions to a different GKZ system. This is due to the fact that, if $F$ is a face of $\A$ there is an isomorphism \cite[Lemma 4.8]{matusevich_homological_2004}
\begin{equation}\label{eq:subfiso}
    H_0(\E+\nu,S_{\A_F})\simeq \C[z_{A\setminus F}]\otimes_\C \M_{\A_F}(\nu)\, ,
\end{equation}
where we recall that $\M_{\A_F}(\nu)$ is the GKZ system defined by the matrix $\A_F$. From the perspective of the solutions of $\M_{\A_F}(\nu)$, this isomorphism simply maps
\begin{equation}
    \phi(1)\mapsto \phi(1)\, .
\end{equation}
Therefore, it is possible to lift the solutions of $\M_{\A_F}(\nu)$ to obtain solutions of $\M_\A(\nu)$. Note that these solutions will automatically satisfy equation~\eqref{eq:pif=02}.

We will now show that a similar story holds even if $F$ is not a face. 
For any subset $F$ of $A$, recall that its toric ring of $\A_F$ is given by
\begin{equation}\label{eq:saf}
    S_{\A_F}\simeq R_\A / \left( I_{\A_F}+  \langle \partial_{\bar{F}}\rangle \right)\,.
\end{equation}
with $R_\A$ the ring of partial derivatives $\partial_I$. Let us consider an $\vert F \vert$-dimensional integer vector $u_F$. It is possible to lift this to an $N$-dimensional vector $u$ by defining $u_I=(u_F)_I$ if $I\in F$ and zero otherwise. Then, any two vectors $u_F$ and $v_F$ satisfying $\A_F u_F=\A_F v_F$ lift to vectors $u$ and $v$ satisfying $\A u= \A v$, since
\begin{equation}
    \A u= \sum_I u_I a_I =\sum_{I\in F}u_I a_I = \A_F u_F\,.
\end{equation}
This implies that 
\begin{equation}
    I_{\A_F}+  \langle \partial_{\bar{F}}\rangle \subseteq I_{\A}+ \langle \partial_{\bar{F}}\rangle\,,
\end{equation}
and applying the third isomorphism theorem results in
\begin{equation}
    R_\A/\left(I_{\A}+  \langle \partial_{\bar{F}}\rangle \right) \simeq \frac{R/\left(I_{\A_F}+  \langle \partial_{\bar{F}}\rangle \right)}{\left(I_{\A}+ \langle \partial_{\bar{F}}\rangle \right)/\left(I_{\A_F}+ \langle \partial_{\bar{F}}\rangle \right)}\, .
\end{equation}
Identifying the left hand side as $S_\A/\langle \partial_{\bar{F}}\rangle$ and relating the right hand side to~\eqref{eq:saf}, this provides us with a surjection
\begin{equation}
    S_{\A_F} \twoheadrightarrow S_\A /\langle \partial_{\bar{F}}\rangle\,.
\end{equation}
By the same arguments as before, this surjection implies that all solutions of $H_0(\E+\nu,S_{\A}/\langle \partial_{\bar{F}}\rangle)$ lift to solutions of $H_0(\E+\nu,S_{\A_F})$. To relate this to the solutions of the actual GKZ system defined by $\A_F$, we note that the Euler operators of $\A_F$ are simply the Euler operators of $\A$ under the map that sends $\partial_I$ to zero for $I$ not in $F$. Therefore, we find that $H_0(\E+\nu, S_{\A_F})$ is isomorphic to the GKZ system defined by $\A_F$ and solutions to $H_0(\E+\nu,S_\A / \langle \partial_{\bar{F}}\rangle)$ lift to solutions to the GKZ system defined by $\A_F$.

The above gives rise to the results included in the first part of section~\ref{ssec:reductionoperators}. In the following, we will explain how the results concerning the reduction operators are obtained in this framework.

\subsection{Reduction operators and their properties.}\label{sap:reductionoperators}

From the discussion above we have seen that, if the parameter $\nu$ is in the $\C$-span for some face $F$, it is possible to obtain a subsystem associated to this face. In general, any surjection of the type
\begin{equation}\label{eq:disurjection}
    S_\A\twoheadrightarrow S_\A/\langle \partial_I\rangle
\end{equation}
can result in a similar subsystem, composed of the solutions of $H_0(\E+\nu,S_\A)$ with $\partial_I f=0$. Note that the existence of such solutions is implied by the existence of a resonant face $F\subseteq A\setminus \{I\}$. In this section, we will show that there exists a different type of submodules for GKZ systems, whose solutions are annihilated by a set of operators we will call reduction operators.

\paragraph{Reduction operators from long exact sequences.}

Equation~\eqref{eq:disurjection} is the end of the exact sequence from equation~\eqref{eq:dIseq}, induced by multiplication with $\partial_I$. Therefore, it is natural to study the induced long exact sequence
\begin{equation}\label{eq:deltaexactseq}
\begin{array}{rll}
    &H_1(\E+\nu,S_\A/\langle \partial_I\rangle)&\xrightarrow{\delta} H_0(\E+\nu+a_I,S_\A)\xrightarrow{\cdot \partial_I} H_0(\E+\nu,S_\A)\\
     \rightarrow &H_0(\E+\nu,S_\A/\langle \partial_I\rangle)&\rightarrow 0\, ,
\end{array}
\end{equation}
where we define $\delta$ as the boundary map arising from the zig-zag lemma. Splicing this long exact sequence in two short exact sequences results in
\begin{equation}\label{eq:imd1ses}
    0\rightarrow \mathrm{im}(\cdot \partial_I)\rightarrow H_0(\E+\nu,S_\A) \rightarrow H_0(\E+\nu,S_\A/\langle \partial_I\rangle)\rightarrow 0\, ,
\end{equation}
and
\begin{equation}\label{eq:imdelses}
    0\rightarrow \mathrm{im}(\delta)\rightarrow H_0(\E+\nu+a_I,S_\A) \xrightarrow{\cdot \partial_I} \mathrm{im}(\cdot \partial_I)\rightarrow 0\, ,
\end{equation}
where $\im(\delta)$ and $\im(\cdot \partial_I)$ are the images of their respective maps. 

The second exact sequence provides us with a submodule of $H_0(\E+\nu+a_I,S_\A)$ by considering its quotient by $\im(\delta)$. The associated solutions will therefore be annihilated by all $Q\in \im(\delta)$. We will call such operators $Q$ \textit{reduction operators} at the parameter $\nu$ in the direction $I$, and we will study their properties in the remaining part of this section.

\paragraph{Number of solutions.}

From combining the exact sequences~\eqref{eq:imd1ses} and~\eqref{eq:imdelses} one can obtain the short exact sequence 
\begin{equation}\label{eq:boundarysequence}
\begin{tikzcd}
    0\arrow[r]& H_0(\E+\nu+a_I,S_\A)/\mathrm{im}(\delta) \arrow[r,"\cdot \partial_I"]& H_0(\E+\nu,S_\A)\arrow[dl]  \\
    & H_0(\E+\nu,S_\A/\langle \partial_I\rangle)\arrow[r] & 0\, .
\end{tikzcd}
\end{equation}
Since this is a short exact sequence of holonomic $\D$-modules, it implies that the rank satisfies
\begin{equation}\label{eq:h0imderanks}
\begin{array}{rll}
   \mathrm{rank}\left(H_0(\E+\nu+a_I,S_\A)/\mathrm{im}(\delta)\right)=& &\mathrm{rank}\left(H_0(\E+\nu,S_\A)\right)\\
   &-&\mathrm{rank}\left(H_0(\E+\nu,S_\A/\langle \partial_I\rangle)\right)\, .
\end{array}
\end{equation}
Therefore, the number of solutions annihilated by the reduction operators at $\nu+a_I$ is determined by the number of solutions annihilated by the partial derivative $\partial_I$ at $\nu$. Correspondingly, if the GKZ system is not rank-jumping the quotient by $\im(\delta)$  is non-trivial if and only if the rank of $H_0(\E+\nu,S_\A/\langle \partial_I\rangle)$ is non-zero.

\paragraph{Partial derivatives map solution spaces.}

Recall from the discussion in section~\ref{sap:dmods} that, since we consider the solutions of these modules around some generic point, we can take the modules we consider to be non-singular. Furthermore, this implies that, when acting on these modules, the solution functor is exact. Applying this observation to the exact sequence~\eqref{eq:boundarysequence} results in
\begin{equation}\label{eq:solimdiseq}
\begin{tikzcd}
        0\arrow[r]& \mathrm{Sol}(H_0(\E+\nu,S_\A/\langle \partial_I \rangle ))\arrow[r] &\mathrm{Sol}(H_0(\E+\nu,S_\A)) \arrow[dl,"\cdot \partial_I"]\\
        &\mathrm{Sol}(H_0(\E+\nu+a_I,S_\A)/\mathrm{im}(\delta))\arrow[r]&0\, ,
\end{tikzcd}
\end{equation}
which provides us with a surjective mapping between the solutions of $H_0(\E+\nu,S_\A)$ and the solutions of $H_0(\E+\nu+a_I,S_\A)/\mathrm{im}(\delta)$ which simply sends $f\mapsto \partial_I f$.

We will now explain how to obtain the image of $\delta$ explicitly, explaining the basis behind the algorithm of section~\ref{ssec:reductionoperators}.

\paragraph{Reduction operators from the zig-zag lemma.}

The map $\delta$ in the sequence~\eqref{eq:dIseq} arises due to the zig-zag lemma, therefore its image can be obtained simply by explicitly performing the diagram chasing.

First, we consider an operator vector $P\in K_1(\E+\nu,S_\A/\langle \partial_I\rangle)$ and require that it satisfies
\begin{equation}\label{eq:tild1p=0}
    \tilde{d}_1(P)=0\, ,
\end{equation}
where $\tilde{d}_1$ is the differential between $K_1(\E+\nu,S_\A/\langle \partial_I\rangle)\rightarrow K_0(\E+\nu,S_\A/\langle \partial_I\rangle)$ . Note that, since this particular $S_\A$ module is not shifted, the differential can be written as
\begin{equation}
    \tilde{d}_1(P)=P\cdot(\A\Theta+\nu)\mod \partial_I\sim0 \in K_0(\E+\nu,S_\A/\langle \partial_I\rangle)\, ,
\end{equation}
where $\cdot$ denotes the vector dot product and we note that working in $K_0(\E+\nu,S_\A/\langle \partial_I\rangle)$ implies that we must impose $\partial_I\sim 0$.

Secondly, we lift $P$ to an element of $K_1(\E+\nu,S_\A)$ and apply the differential $d_1:K_0(\E+\nu,S_\A)\rightarrow K_0(\E+\nu,S_\A)$ to $P$. In order to do this, we again calculate $P\cdot (\A\Theta+\nu)$, however, since we are now considering elements of $ K_0(\E+\nu,S_\A)$ we must no longer set $\partial_I ~0$. This recovers the second step of the algorithm.

Finally, this procedure guarantees that 
\begin{equation}
    P\cdot (\A \Theta+\nu)=Q\partial_I \in K_0(\E+\nu,S_\A)
\end{equation}
for some operator $Q$. The action of $\delta$ on $P$ is then defined as
\begin{equation}
    \delta(P)=Q
\end{equation}
resulting in the reduction operator $Q$. In practice, it is enough to find the generators of $\im(\delta)$, since these will be the relevant operators when discussing solutions. 

In order to obtain these generators, one can start by considering the different solutions to equation~\eqref{eq:tild1p=0}. Noting that since we want to consider elements of
\begin{equation}
    H_1(\E+\nu,S_\A/\langle \partial_I\rangle)=\ker(\tilde{d}_1)/\im(\tilde{d}_2)
\end{equation}
we should ignore relations due to operators in $\im(\tilde{d}_2)$. In practice, this requires us to use the toric operators of $\A$ that have a term proportional to $\partial_I$. From these toric operators it is often clear how many such solutions are possible and from these solutions one can then obtain a full basis of generators of $\im(\delta)$.

\paragraph{Homogeneity and $\partial_I Q=0$.}

Although the algorithm above works for any $P \in K_1(\E+\nu,S_\A/\langle \partial_I\rangle)$, there are two particular classes of $P$ satisfying some additional useful properties. Firstly, it is often useful to consider only those reduction operators that are homogeneous with respect to the $\A$ grading. This is possible since $K_\bullet (\E+\nu,S_\A)$ and $K_\bullet (\E+\nu,S_\A/\langle \partial_I \rangle)$, as well as their differentials, inherit the $\A$ grading \cite[Lemma 4.3]{matusevich_homological_2004}. From this, one finds that by decomposing 
\begin{equation}
    P=\sum_{\alpha} P_\alpha
\end{equation}
into its homogeneous elements $P_\alpha$, each $P_\alpha$ must satisfy
\begin{equation}
    \tilde{d}_1(P_\alpha)=0
\end{equation}
separately. Therefore it is possible to obtain a reduction operator
\begin{equation}
    Q_\alpha\coloneqq \delta(P_\alpha)\, ,
\end{equation}
for each $P_\alpha$. Furthermore, since $\delta$ is compatible with the grading, $Q_\alpha$ is homogeneous. This allows us to always find a homogeneous set of generators for $\im(\delta)$. Note that this homogeneity implies that
\begin{equation}
    [\A \theta,Q_\alpha]=-\alpha Q
\end{equation}
since $\E_J$ act as the grading operators.

Secondly, we will show that it is always possible to obtain generators $Q$ that satisfy
\begin{equation}
    \partial_I Q=0 \in H_0(\E+\nu+a_I,S_\A)\, ,
\end{equation}
where the derivative $\partial_I$ acts on everything to the right. To show this we need that it is possible to obtain operator vectors $P$ independent of $z_I$. To see this, note that
\begin{equation}
    \D\otimes_{R} S_\A/\langle\partial_I\rangle \simeq \C[z_I]\otimes_\C( \D_{A\setminus \{I\}}\otimes_{R_{A\setminus \{I\}}}S_\A/\langle \partial_I \rangle)\, ,
\end{equation}
where we recall that $\D_{A\setminus\{I\}}$ and $R_{A\setminus\{I\}}$ denote the Weyl algebra and polynomial ring in the variables indexed by $A\setminus \{I\}$. Furthermore, writing $\E^{A\setminus\{I]}$ for the Euler operators with $\partial_I$ set to zero, it is possible to adapt the proof of \cite[Lemma 4.8]{matusevich_homological_2004} to arbitrary subsets $F\subset A$. This allows us to decompose
\begin{equation}
    K_\bullet (\E+\nu,S_\A/\langle \partial_I \rangle)\simeq \C[z_I]\otimes_{\C} K^{A\setminus\{I\}}_\bullet (\E^{A\setminus\{I\}}+\nu,S_\A/\langle \partial_I \rangle)\, ,
\end{equation}
where we have defined $K^{A\setminus\{I\}}_\bullet$ as the Euler-Koszul complex over the ring $R_{A\setminus \{I\}}$ and $\E^{A\setminus\{I\}}$ as the associated reduction operators.\footnote{Note that, while the complex $K^{A\setminus\{I\}}(\E^{A\setminus\{I\}}+\nu)$ is exactly the complex obtained from the matrix $\A_{A\setminus\{I\}}$, the rings $S_\A/\langle \partial_I\rangle$ and $S_{\A_{A\setminus\{I\}}}$ are not isomorphic in general. Therefore, this construction does not immediately result in a map between ordinary GKZ systems.} The differential of this complex acts trivially on $\C[z_I]$, therefore the generators $P$ can be chosen to be independent of $z_I$.

A consequence of $(\partial_I P)=0$ is that its reduction operator $Q$ satisfies
\begin{equation}\label{eq:partialiqi}
    (\partial_I Q)=P\cdot a_I\, ,
\end{equation}
which follows from the observation that
\begin{equation}
    P\cdot \A\Theta=P_1+(P\cdot a_I)z_I\partial_I
\end{equation}
with $P_1$ independent of $z_I$. Since $Q\partial_I = P\cdot( \A \Theta+\nu)$, equation~\eqref{eq:partialiqi} follows. In order to use this to show that $\partial_I Q=0$, simply commute
\begin{equation}
    \partial_I Q=Q\partial_I+(\partial_I Q) = P\cdot (\A\Theta+\nu)+P\cdot a_I\, ,
\end{equation}
where in the second equality we have applied the definition of the reduction operator. Since all components of the vector $A\Theta+\nu+a_I$ are zero in $H_0(\E+\nu+a_I,S_\A)$, we find that $\partial_I Q$ vanishes.

\bibliographystyle{utphys}
\providecommand{\href}[2]{#2}\begingroup\raggedright\endgroup

\end{document}